\documentclass[11pt,a4paper]{article}

\pdfoutput=1

\usepackage{graphicx}
\usepackage{comment}
\usepackage{jheppub}
\newcommand{\ee}{\end{equation}}
\newcommand{\br}{\begin{eqnarray}}
\newcommand{\bea}{\begin{eqnarray}}
\newcommand{\eea}{\end{eqnarray}}
\newcommand{\er}{\end{eqnarray}}
\newcommand{\ba}{\begin{array}}
\newcommand{\ea}{\end{array}}

\newcommand{\bi}{\begin{itemize}}
\newcommand{\ei}{\end{itemize}}
\newcommand{\bn}{\begin{enumerate}}
\newcommand{\en}{\end{enumerate}}
\newcommand{\bc}{\begin{center}}
\newcommand{\ec}{\end{center}}

\newcommand{\beq}{\begin{equation}}
\newcommand{\eeq}{\end{equation}}

\newcommand{\MP}{{M_{\rm P}}}

\usepackage{subfig}
\usepackage{float}
\usepackage{bigints}

\title{\centering  Multiple point criticality principle\\ and Coleman-Weinberg inflation}

\author[a]{Antonio Racioppi}
\affiliation[a]{National Institute of Chemical Physics and Biophysics, R\"avala 10, 10143 Tallinn, Estonia}

\author[b]{J\"urgen Rajasalu}
\affiliation[b]{Tallinn University of Technology, Akadeemia tee 23, 12618 Tallinn, Estonia}

\author[b]{Kaspar Selke}

\emailAdd{antonio.racioppi@kbfi.ee}
\emailAdd{jyrgen.rajasalu@gmail.com}
\emailAdd{selke.kaspar@gmail.com}

\abstract{We apply the multiple point criticality principle to inflationary model building and study Coleman-Weinberg inflation when the scalar potential is quadratic in the logarithmic correction. We analyze also the impact of a non-minimal coupling to gravity under two possible gravity formulation: metric or Palatini. We compare the eventual compatibility of the results with the final data release of the Planck mission.}

\keywords{Inflation, non-minimal coupling, Palatini, multiple point criticality principle}

\begin{document}
\maketitle

\section{Introduction} \label{sec:Introduction}
During the initial moments of its life, the Universe underwent a period of exponential expansion known as cosmic inflation~\cite{Starobinsky:1980te,Guth:1980zm,Linde:1981mu,Albrecht:1982wi}. 
The theory of cosmic inflation has the merit of providing simultaneously a solution to issues like the flatness and horizon problems of the Universe and a way to generate primordial inhomogeneities, whose power spectrum has been tested in several experiments~\cite{Array:2015xqh,Planck2018:inflation}. 
Already the very first papers on inflation \cite{Linde:1981mu,Albrecht:1982wi}
considered radiatively induced inflaton potentials {\it \`{a}~la} Coleman-Weinberg (CW)~\cite{Coleman:1973jx}.  Such an idea has been extensively studied in the context of grand unified theories, in $U(1)_{B-L}$ extension of the
SM (e.g. \cite{Okada:2019bqa,Biondini:2020xcj,Borah:2020wyc} and refs. therein), or with scalar extensions in \cite{Kannike:2014mia}.  It has been demonstrated that radiative corrections to inflationary potentials may play a relevant role \cite{Kannike:2014mia,Marzola:2015xbh,Marzola:2016xgb,Dimopoulos:2017xox}, dynamically generating the Planck scale \cite{Salvio:2014soa,Kannike:2015apa,Kannike:2015kda}, leading to linear inflation predictions in presence of a non-minimal coupling to gravity \cite{Kannike:2015kda,Rinaldi:2015yoa,Barrie:2016rnv,Artymowski:2016dlz,Racioppi:2017spw,Racioppi:2018zoy}, or predicting super-heavy dark matter \cite{Farzinnia:2015fka,Kannike:2016jfs}.
However most of the previous studies are assuming that the running of the inflaton self-quartic coupling is essentially linear in the radiatively generated logarithmic corrections. In this article we instead assume that the leading order is quadratic in such corrections. The motivation relies in the multiple point criticality principle (MPCP) (e.g. \cite{McDowall:2019knq} and refs. therein), which states that nature chooses the Higgs potential parameters so that different phases of electroweak symmetry breaking may coexist. Operatively, it means that the Higgs potential possesses multiple (nearly) degenerate minima. Such a principle was introduced already in 1995 \cite{Froggatt:1995rt} and used to predict the measured Higgs mass with a surprisingly good accuracy. The applications of the MPCP have been multiple (e.g. \cite{McDowall:2019knq,Kannike:2020qtw} and refs therein). For what concerns inflation, most of the efforts were focused on the analyses of different versions of SM Higgs inflation (e.g. \cite{Shaposhnikov:2020geh} and refs. therein) where the MPCP is usually slightly broken (see also \cite{Hamada:2014xka,Kawana:2015tka}), or Agravity-like theories (e.g. \cite{Salvio:2014soa,Kannike:2015apa} and refs. therein) where the MPCP is indeed exact. Moreover, the MPCP is also a powerful tool that allows to tune of the cosmological constant in our vacuum to zero. When the inflaton potential is quartic and subject to (not self-induced) radiative corrections, it inevitably develops a vacuum expectation value different from zero which generates a non-null cosmological constant (e.g \cite{Kannike:2014mia,Salvio:2014soa} and refs there in). Usually the issue is solved by adding by hand an opposite constant, so that the net effect is a null constant in the vacuum. A more elegant way is to impose the MPCP and have a quartic potential with the self-quartic coupling that runs quadratically in the logarithmic correction (e.g. \cite{Salvio:2014soa} and refs. therein). In this article we study this kind of scenario, model independently and assuming the inflaton not to be necessarily the Higgs boson, but a generic scalar.
 
When studying radiatively corrected inflaton potentials, is almost inevitable to discuss
also non-minimal couplings to gravity, which naturally arise from quantum corrections in a curved space-time \cite{Birrell:1982ix}.
 In particular, this happens when the SM Higgs scalar is the inflaton field \cite{Bezrukov:2007ep}. In this case we have a non-minimal coupling to gravity of the type $\xi \phi^2 R$, where $\phi$ is the inflaton field, $R$ the Ricci scalar and $\xi$ a coupling constant. This kind of models have been studied in a large number of works over the past decades (in e.g. \cite{Futamase:1987ua,Salopek:1988qh,Fakir:1990eg,Amendola:1990nn,Kaiser:1994vs,Bezrukov:2007ep,Bauer:2008zj,Park:2008hz,Linde:2011nh,Kaiser:2013sna,Kallosh:2013maa,Kallosh:2013daa,Kallosh:2013tua,Galante:2014ifa,Chiba:2014sva,Boubekeur:2015xza,Pieroni:2015cma,Jarv:2016sow,Salvio:2017xul,Karam:2017rpw,Bostan:2018evz,Almeida:2018pir,Cheng:2018axr,Tang:2018mhn,SravanKumar:2018tgk,Kubo:2018kho,Canko:2019mud,Okada:2019opp,Karam:2018mft,Kubo:2020fdd} and refs. therein). In this article we are going to add such non-minimal couplings to models of CW inflation and MPCP. 
The presence of non-minimal couplings to gravity requires then a discussion about the gravitational degrees of freedom. In the usual metric formulation of gravity the independent variables are the metric and its first derivatives, while in the Palatini formulation \cite{Bauer:2008zj} the independent variables are the metric and the connection. Using the Einstein-Hilbert Lagrangian, the two formalisms predict the same equations of motion and therefore describe equivalent physical theories. 
However, with non-minimal couplings between gravity and matter, such equivalence is lost and the two formulations describe different gravity theories \cite{Bauer:2008zj} and lead to different phenomenological results, as recently investigated in (e.g. \cite{Koivisto:2005yc,Tamanini:2010uq,Bauer:2010jg,Rasanen:2017ivk,Tenkanen:2017jih,Racioppi:2017spw,Markkanen:2017tun,Jarv:2017azx,Racioppi:2018zoy,Kannike:2018zwn,Enckell:2018kkc,Enckell:2018hmo,Rasanen:2018ihz,Bostan:2019uvv,Bostan:2019wsd,Carrilho:2018ffi,Almeida:2018oid,Takahashi:2018brt,Tenkanen:2019jiq,Tenkanen:2019xzn,Tenkanen:2019wsd,Kozak:2018vlp,Antoniadis:2018yfq,Antoniadis:2018ywb,Gialamas:2019nly,Racioppi:2019jsp,Rubio:2019ypq,Lloyd-Stubbs:2020pvx,Das:2020kff,McDonald:2020lpz,Shaposhnikov:2020fdv,Enckell:2020lvn,Jarv:2020qqm,Gialamas:2020snr,Karam:2020rpa,Gialamas:2020vto,Karam:2021wzz,Karam:2021sno,Gialamas:2021enw,Annala:2021zdt} and refs. therein). 

The aim of this work is to combine inflation and the MPCP by studying CW inflation when the dominant loop contribution is a squared logarithm, with and without a non-minimal coupling to gravity and considering two possible gravity formulation (metric or Palatini). 
The article is organized as follows.
In section \ref{sec:CW} we establish the setup for the MPCP and the CW potential and study the corresponding inflationary phenomenology in Einsteinian gravity. Then in section \ref{sec:xiCW} we study the same setup but in presence of a non-minimal coupling to gravity and under two different gravity formulations: metric or Palatini. Finally in section \ref{sec:conclusions} we present our conclusions.

\section{Coleman-Weinberg inflation and multiple-point criticality principle } \label{sec:CW}
Consider the following action describing Einsteinian gravity plus an inflaton scalar $\chi$
\begin{equation}
S = \!\! \int \!\! d^4x \sqrt{-g}\left(-\frac{M_P^2}{2} R + \frac{(\partial \chi)^2}{2}  - U(\chi) \right) ,
\label{eq:L:classic}
\end{equation}
where $M_P$ is the reduced Planck mass, $R$ is the Ricci scalar and $U(\chi)$ is the effective potential of the inflaton, that we assume to behave as
\begin{equation}
  U(\chi) = \lambda_{\rm eff} (\chi) \chi^4 \, .
  \label{eq:U:v1}
\end{equation}
The quartic coupling pre-factor in eq.~\eqref{eq:U:v1} is subject to quantum corrections.
We study now the minimization of the scalar potential\footnote{A complete discussion was already presented in \cite{Racioppi:2017spw,Gialamas:2020snr}, however for the sake of clarity we repeat the relevant details.}. The general equation for the stationary points of the scalar potential in eq.~\eqref{eq:U:v1} is
\begin{equation}
 \Big[ 4 \lambda_{\rm eff} (\chi)  + \beta_{\rm eff} (\chi) \Big] \chi^3   =0 \, , \label{eq:min:RGE}
\end{equation}
where $\beta_{\rm eff} (\mu)= \mu \frac{\partial}{\partial \mu}\lambda_{\rm eff} (\mu)$ is the beta-function of the quartic coupling $\lambda_{\rm eff} (\mu)$. One trivial solution of eq. \eqref{eq:min:RGE} is $\chi=0$. Additional solutions may appear if it is possible to find  a $\chi=M$ so that
\begin{equation}
 4 \lambda_{\rm eff} (M) + \beta_{\rm eff} (M)  =0 \, . \label{eq:min:RGE:2}
\end{equation}
Such an equation has three possible solutions:
\begin{eqnarray}
\text{a)} & \beta_{\rm eff} (M)=\lambda_{\rm eff} (M)=0 , \label{eq:RGE:MPCP}\quad\\
\text{b)} & \beta_{\rm eff} (M) > 0, \ \lambda_{\rm eff} (M)<0 ,\label{eq:RGE:bound:cond}\\
\text{c)} & \beta_{\rm eff} (M) < 0, \ \lambda_{\rm eff} (M)>0  .
\end{eqnarray}
 Here enters the MPCP. By applying it, we require that both $\chi=0$ and $\chi=M$ are degenerate minima of the potential, leaving option a) as the only viable solution. Without knowing the details of the whole theory and its particle content, we can  model-independently write $\lambda_{\rm eff}$ as a Taylor expansion around the scale $M$:
\begin{equation}
\lambda_{\rm eff}(\chi) = \lambda_0(M) + \lambda_1 (M) \ln\frac{\chi }{M}+
                  \frac{1}{2!} \lambda_2 (M) \ln^2\frac{\chi }{M} +
                     \frac{1}{3!} \lambda_3 (M)  \ln^3\frac{\chi }{M} + \cdots ,
  \label{eq:lambdaTaylor}
\end{equation}
where $\lambda_i (\mu)$ is the $i$-th derivative of $\lambda(\mu)$ with respect to $t=\ln \mu$ and we assumed without loss of generality that $\chi>0$. Note that $\lambda_1=\beta_{\rm eff}$. The MPCP requires $\lambda_0(M)=\lambda_1 (M)=0$. Such a behaviour is analogous to the one of the self-quartic coupling of the SM Higgs boson (e.g. \cite{Shaposhnikov:2020geh,Shaposhnikov:2020fdv} and refs. therein), after which the MPCP was indeed proposed (e.g. \cite{McDowall:2019knq} and refs. therein). Therefore, inspired by it, but without restricting $\chi$ to be necessarily the SM Higgs boson, we assume that during inflation the dominant contribution to $\lambda_{\rm eff}(\chi)$ in eq. \eqref{eq:lambdaTaylor} comes from the $\lambda_2$ term. Doing this, we are left with
\begin{equation}
    \lambda_{\rm eff}(\chi) \simeq \alpha \ln^2\left(\frac{\chi}{M}\right) \, ,
    \label{eq:lambda:run}
\end{equation}
where $\alpha=\lambda_2/2$ is treated as a free parameter of our model.
Therefore now the inflaton potential reads
\begin{equation}
    U(\chi) = \alpha \ln^2\! \left(\frac{\chi }{M}\right) \chi ^4 \, .
    \label{eq:U:final}
\end{equation}
Assuming slow-roll, the inflationary dynamics is described by the usual slow-roll parameters 
\beq
\epsilon \equiv \frac{1}{2}M_{\rm P}^2 \left(\frac{1}{U}\frac{{\rm d}U}{{\rm d}\chi}\right)^2 \,, \quad
\eta \equiv M_{\rm P}^2 \frac{1}{U}\frac{{\rm d}^2U}{{\rm d}\chi^2} \,.
\ee
Inflation happens when $\epsilon \ll 1$. The corresponding expansion of the Universe is measured in number of $e$-folds
\beq
N_e = \frac{1}{M_{\rm P}^2} \int_{\chi_f}^{\chi_N} {\rm d}\chi \, U \left(\frac{{\rm d}U}{{\rm d} \chi}\right)^{-1},
\label{eq:Ne}
\ee
where the field value at the end of inflation, $\chi_f$, is defined via $\epsilon(\chi_f) = 1$.  %
\begin{figure}[t]
    \centering
    \includegraphics[width=0.49\textwidth]{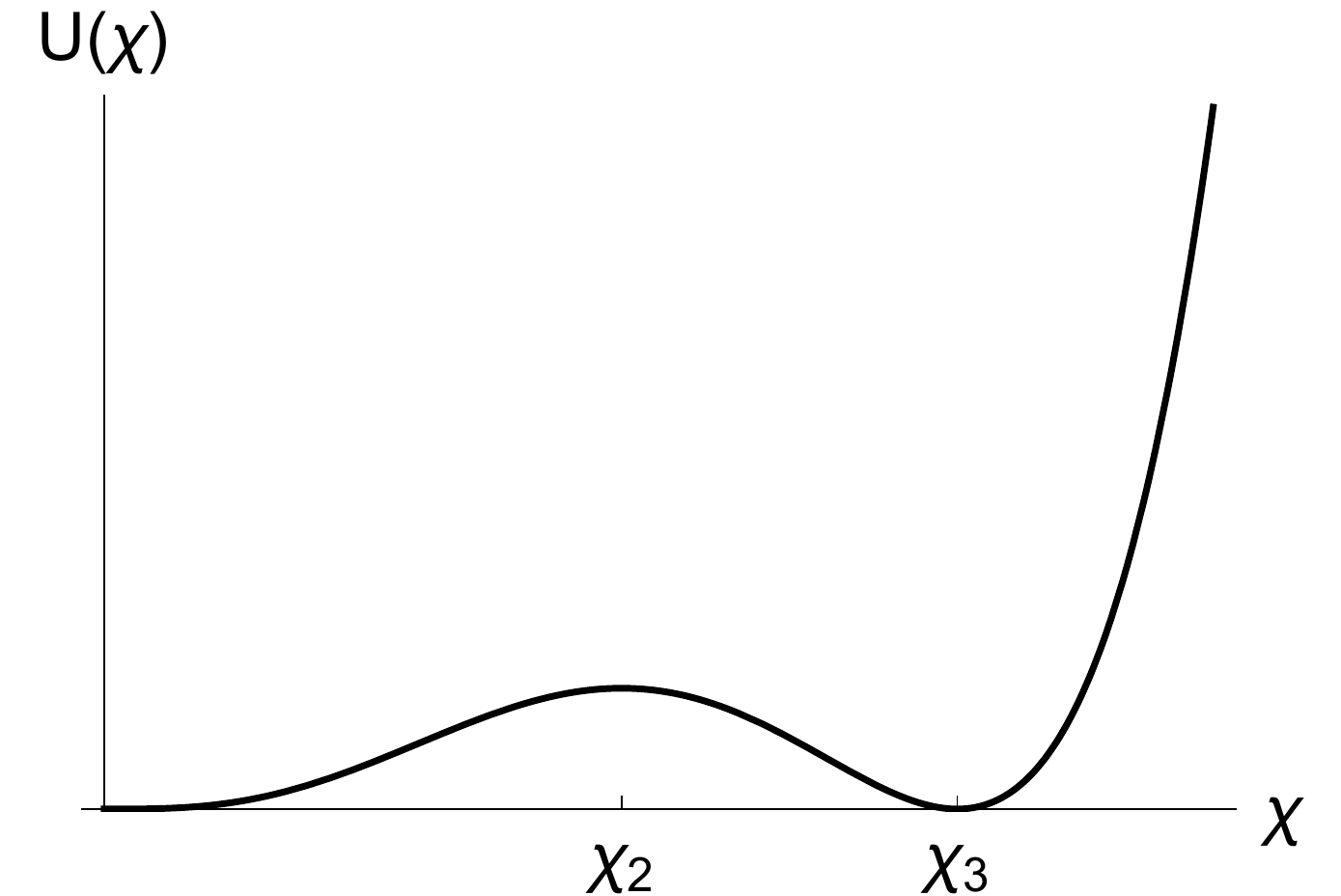}
     \includegraphics[width=0.49\textwidth]{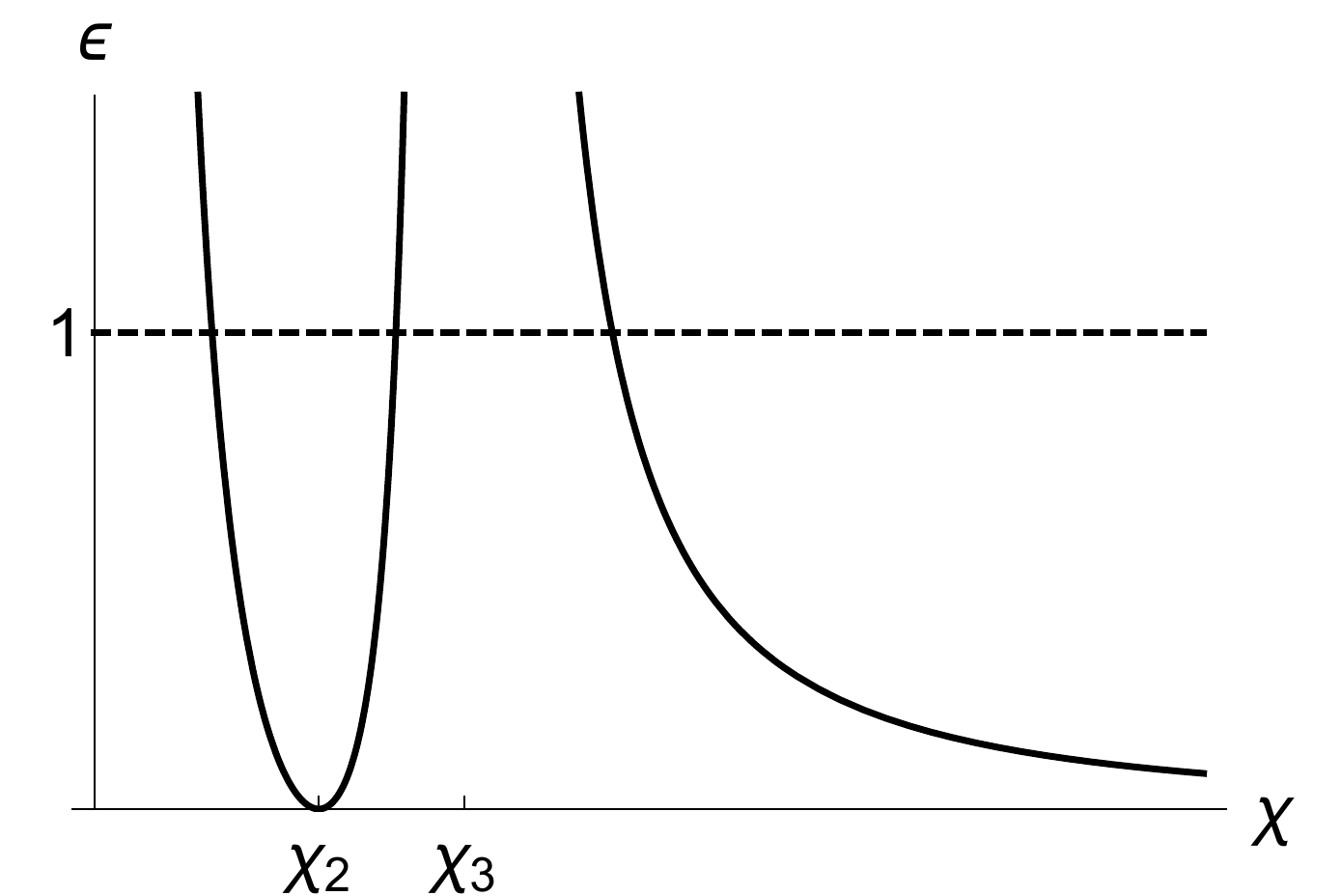}
    \caption{Reference plot for $U(\chi)$ (left) and for $\epsilon$ (right) in function of $\chi$.}
    \label{fig:V_and_eps_CW}
\end{figure}
\begin{figure}[t]
    \subfloat[]{\includegraphics[width=0.46\textwidth]{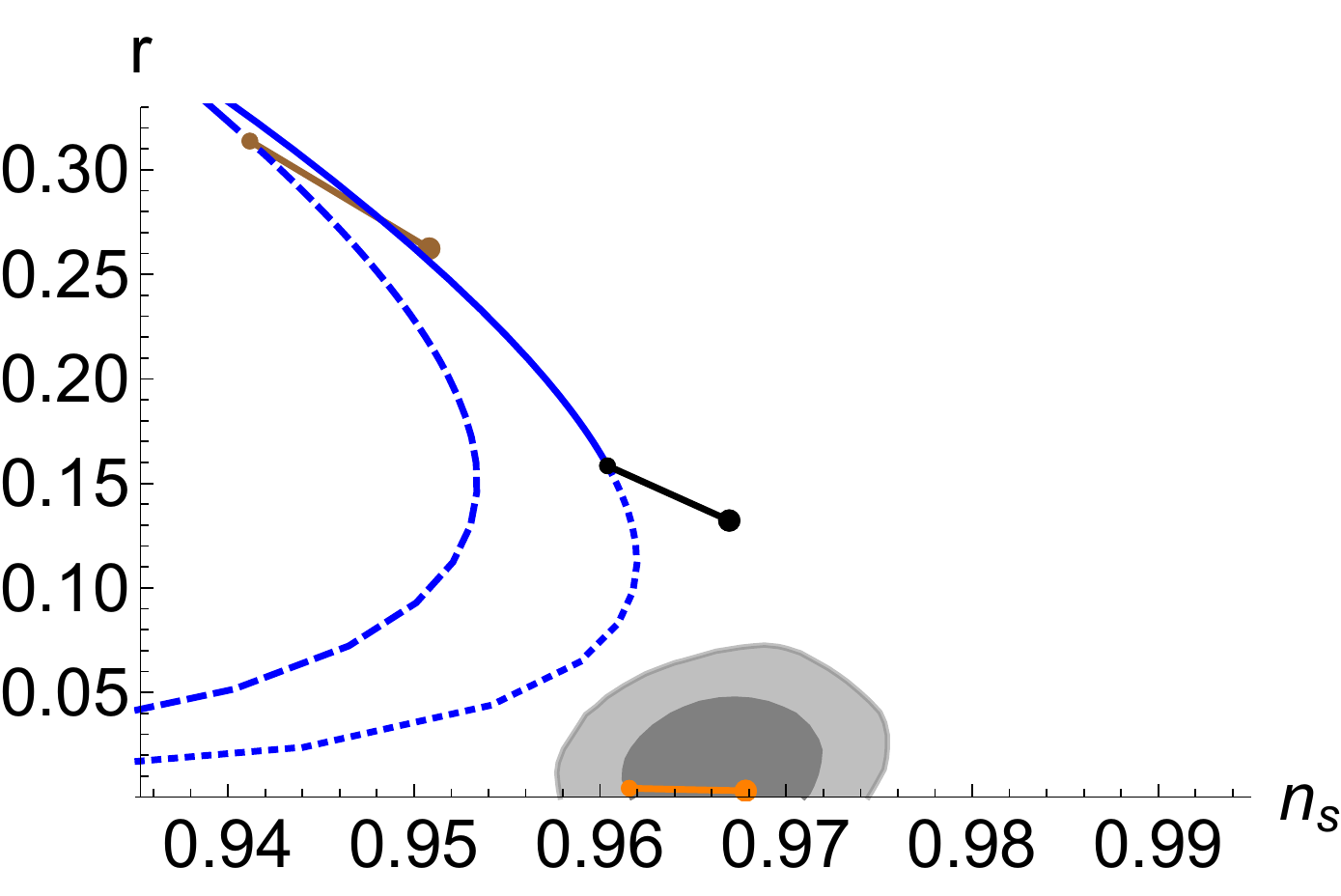}}%
    \subfloat[]{\includegraphics[width=0.46\textwidth]{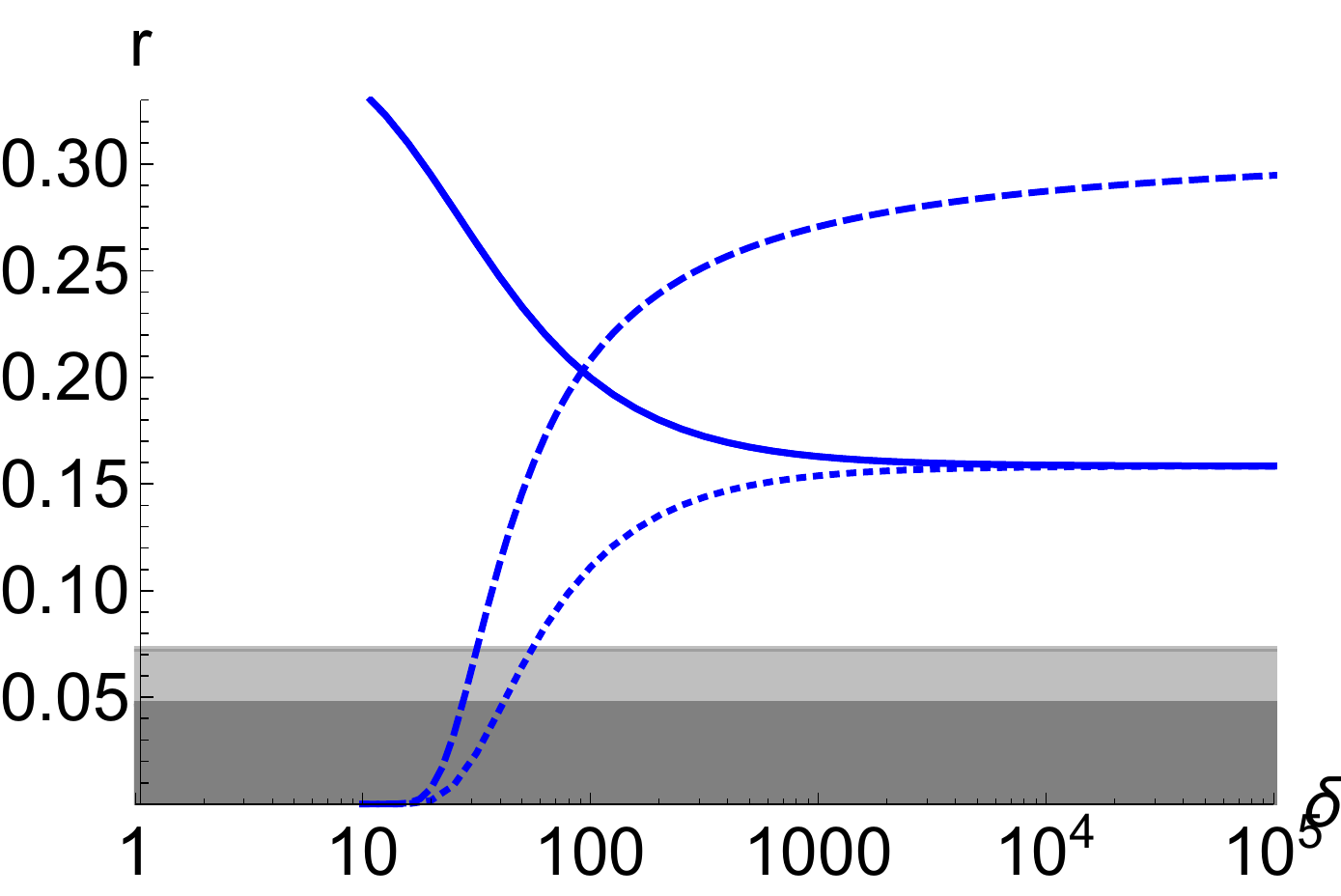}}%
    
    \subfloat[]{\includegraphics[width=0.46\textwidth]{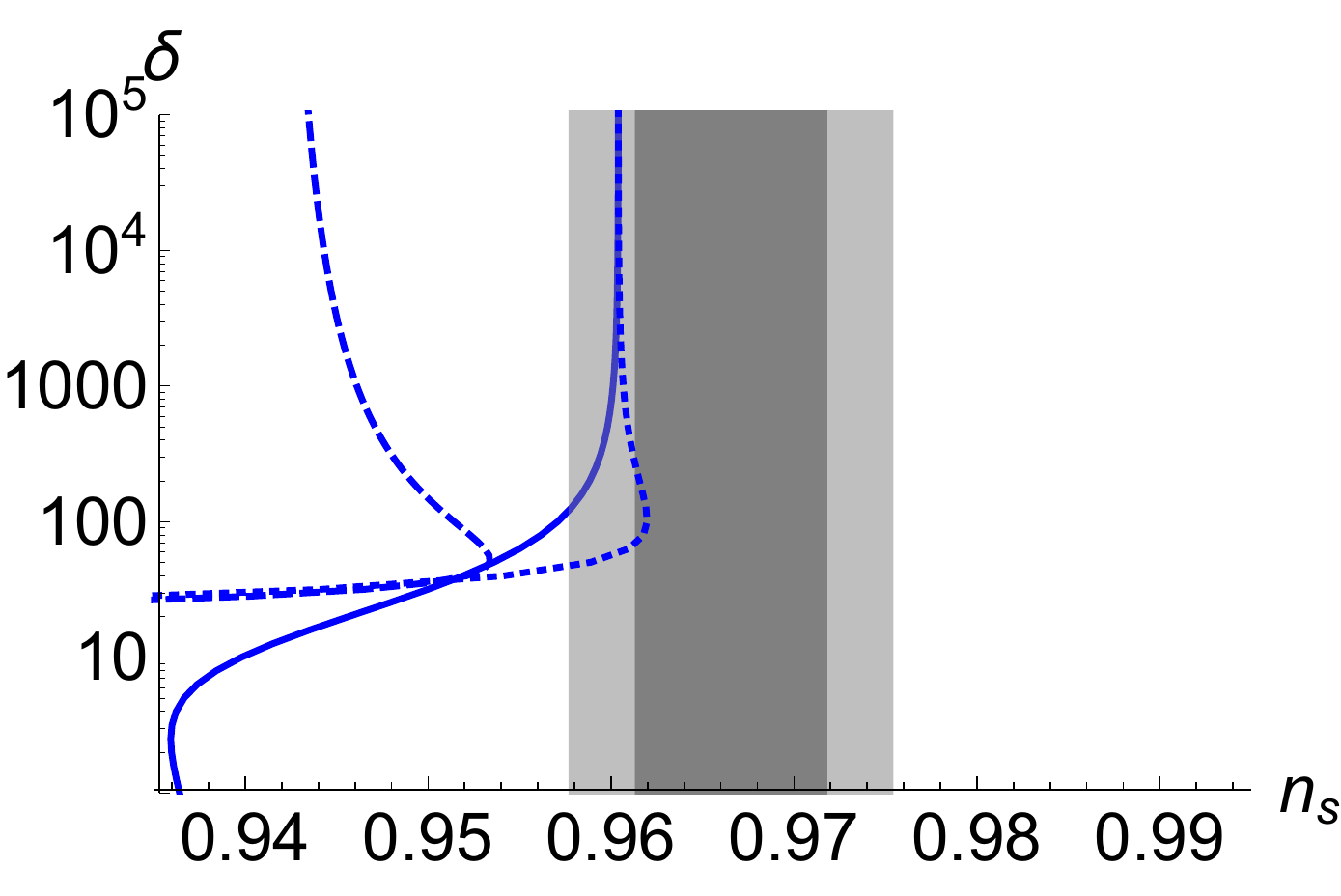}}%
    \subfloat[]{\includegraphics[width=0.46\textwidth]{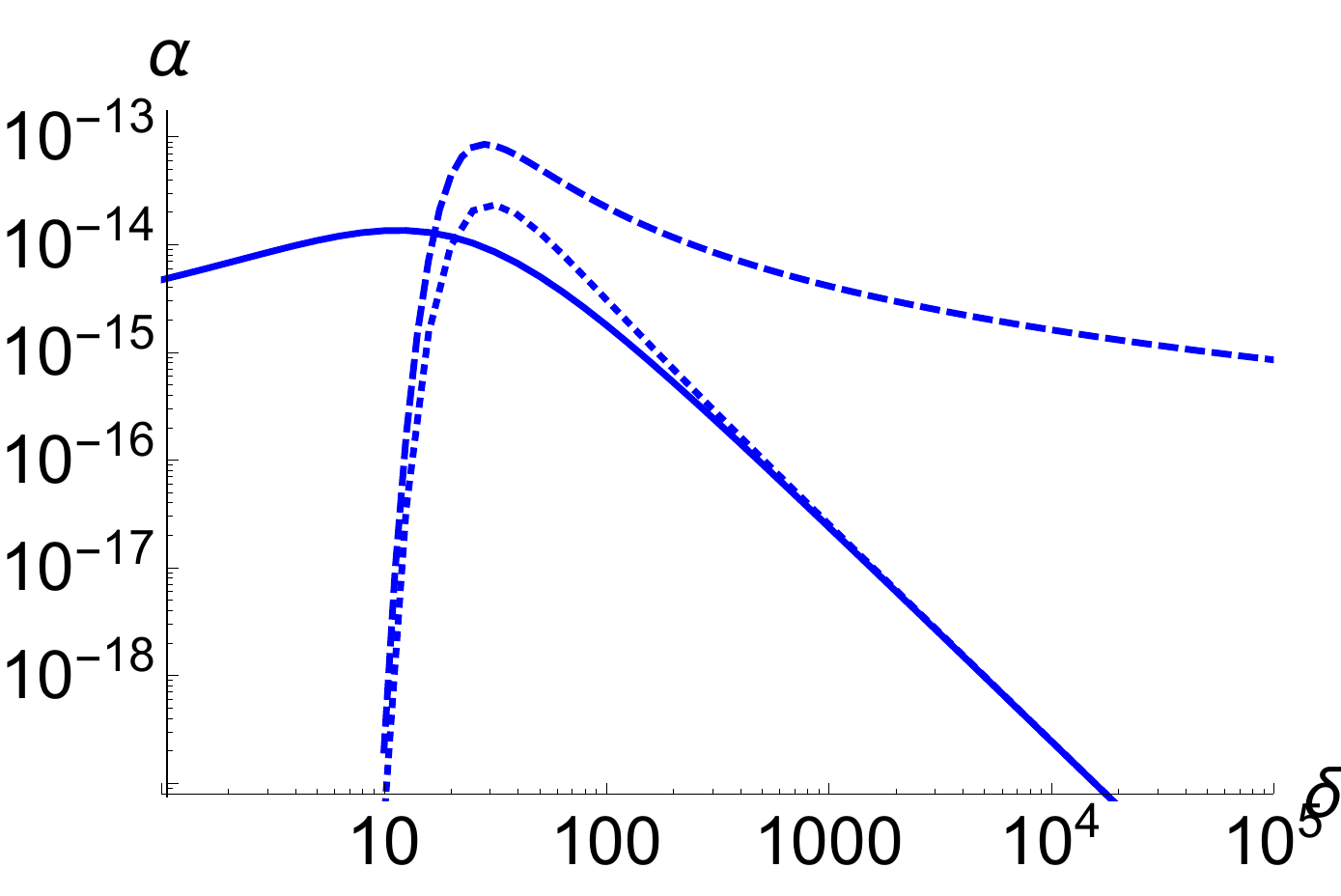}}%
\caption{$r$ vs. $n_s$ (a), $r$ vs. $\delta$ (b), $\delta$ vs. $n_s$ (c), and $\alpha$ vs. $\delta$ (d) with $N_e = 50$ $e$-folds for the CW potential \eqref{eq:U:final}. Blue dashed line represents inflation in region 1), blue dotted line in region 2) and blue continuous in region 3). For reference we plot the lines for quartic (brown), quadratic (black) and $R^2$ (orange) inflation for $N_e \in [50,60]$. The gray areas represent the 1,2$\sigma$ allowed regions from Planck 2018 data~\cite{Planck2018:inflation}.}
\label{fig:CW:50}
\begin{center}
    \includegraphics[width=0.5\textwidth]{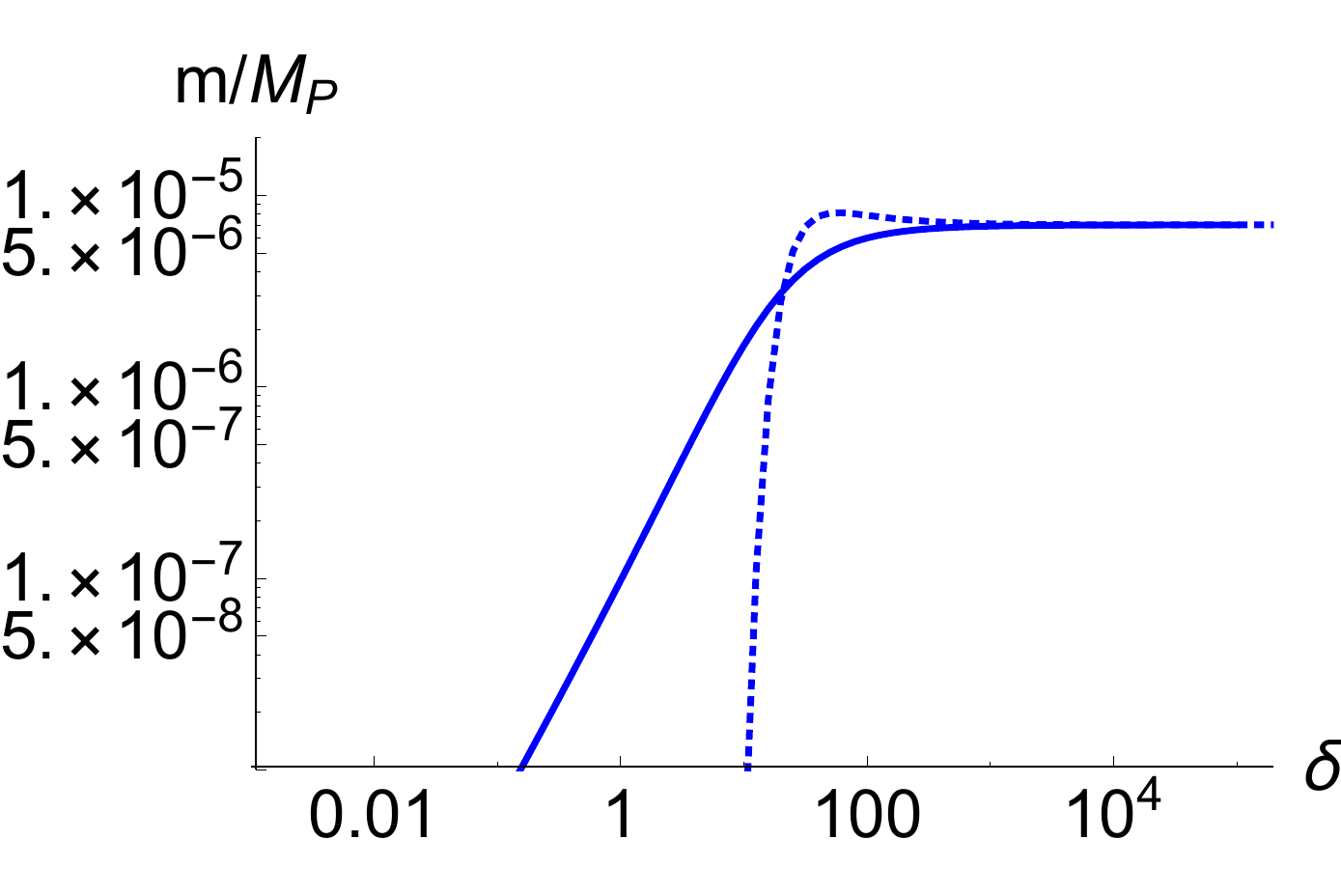}
\end{center}
\caption{Inflaton mass $m$ vs. $\delta$ with $N_e = 50$ $e$-folds for the CW potential \eqref{eq:U:final}. Blue dotted line represents inflation in region 2) and blue continuous in region 3).}
\label{fig:m:CW:50}
\end{figure}

\begin{figure}[t]
    \subfloat[]{\includegraphics[width=0.46\textwidth]{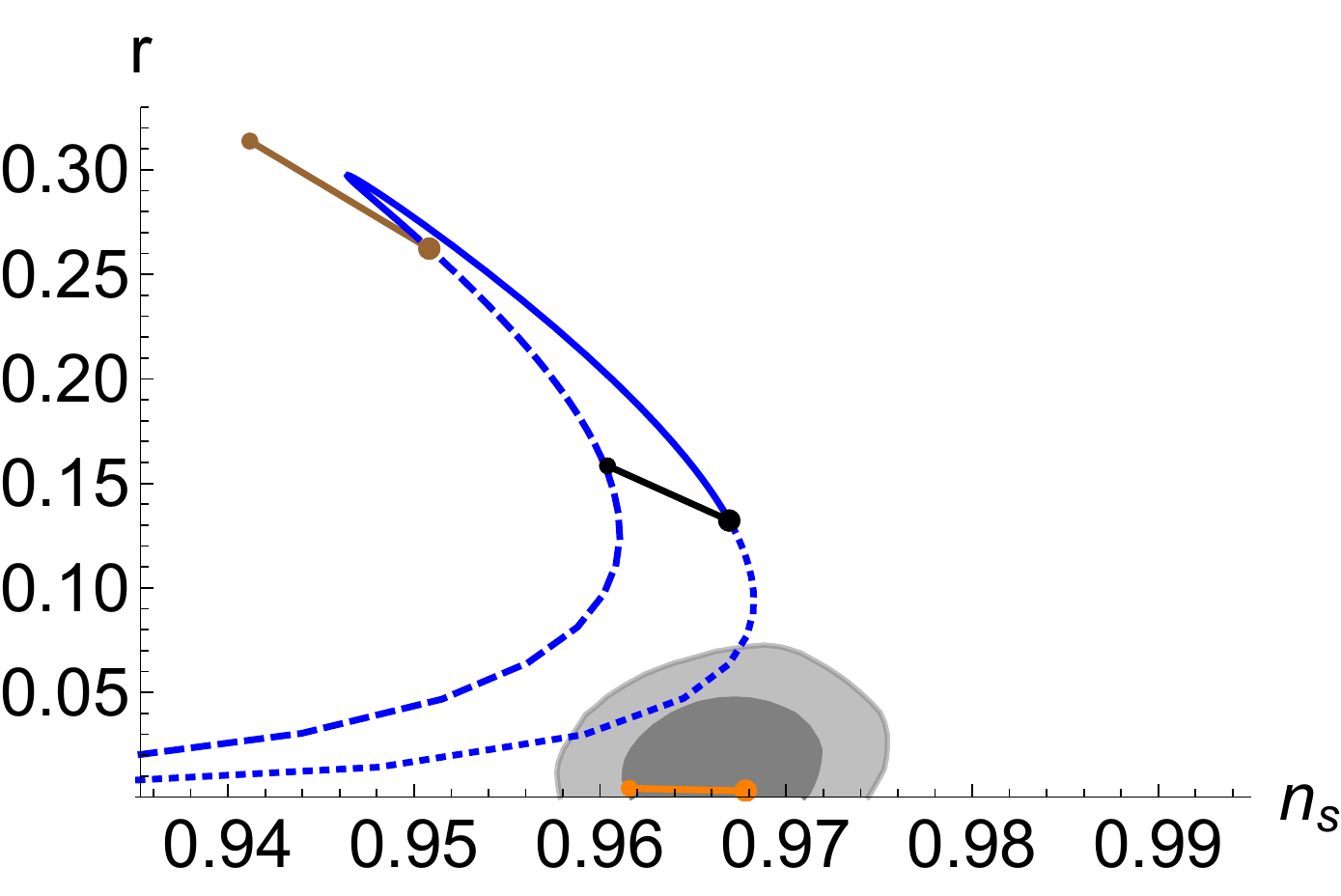}}%
    \subfloat[]{\includegraphics[width=0.46\textwidth]{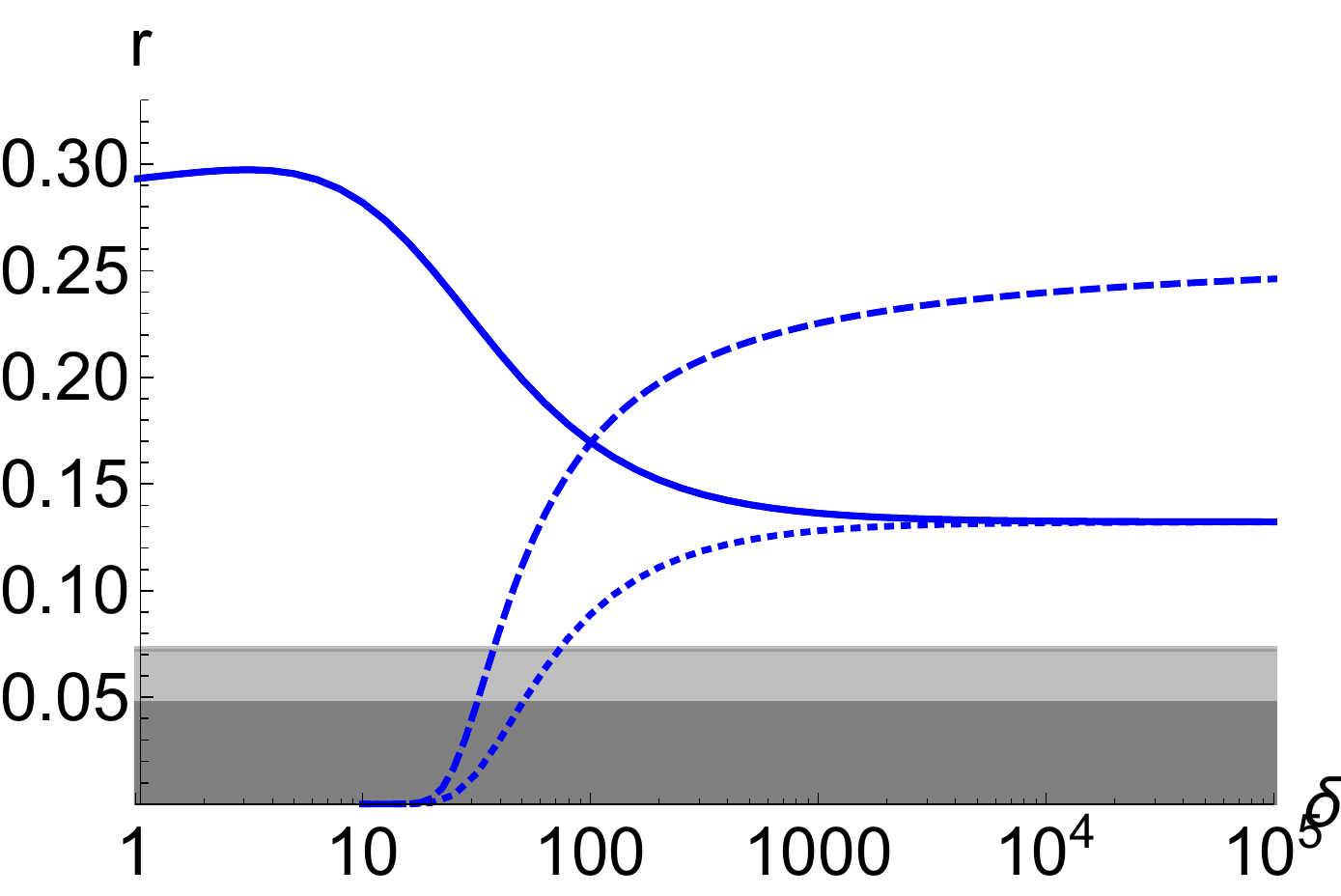}}%
    
    \subfloat[]{\includegraphics[width=0.46\textwidth]{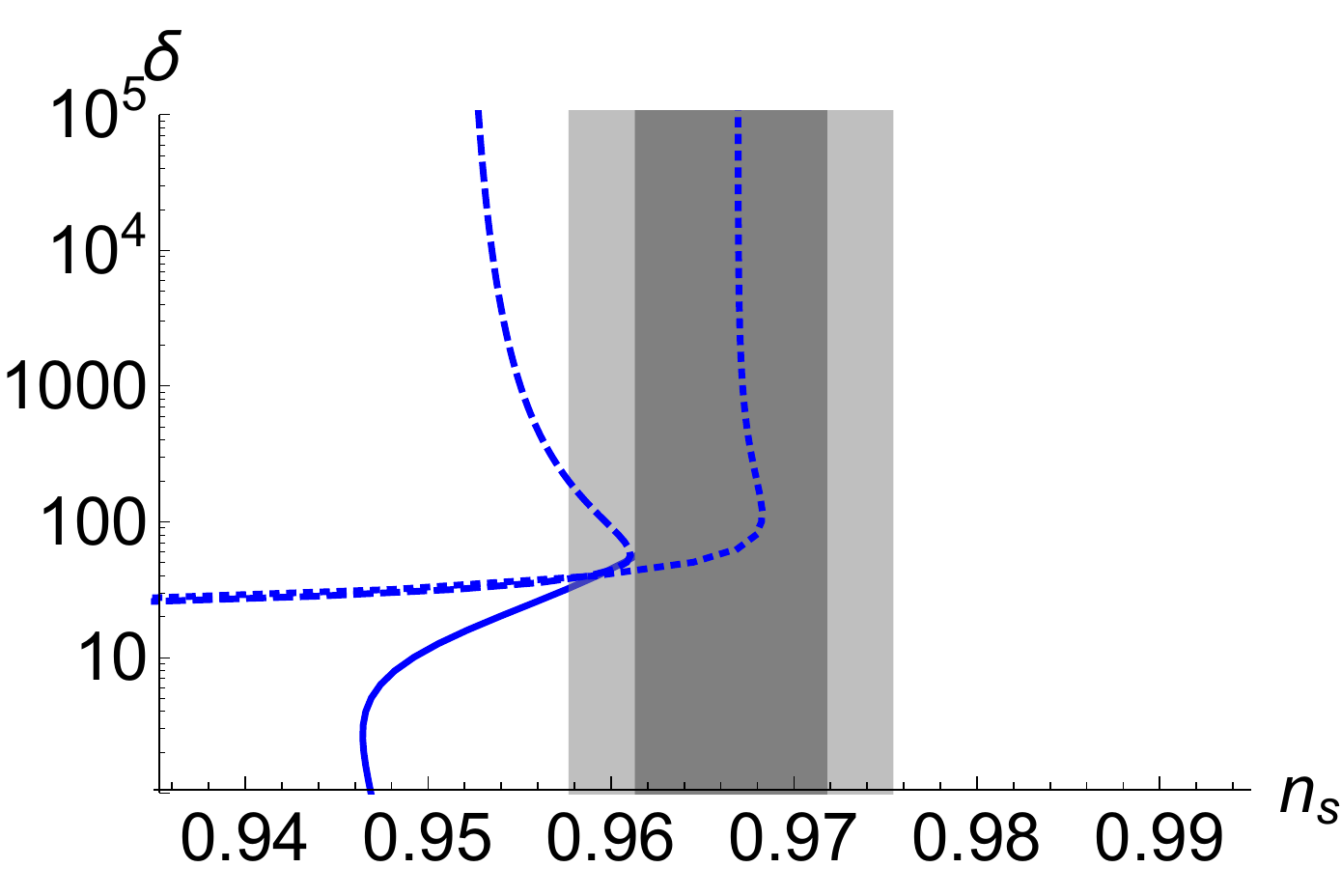}}%
    \subfloat[]{\includegraphics[width=0.46\textwidth]{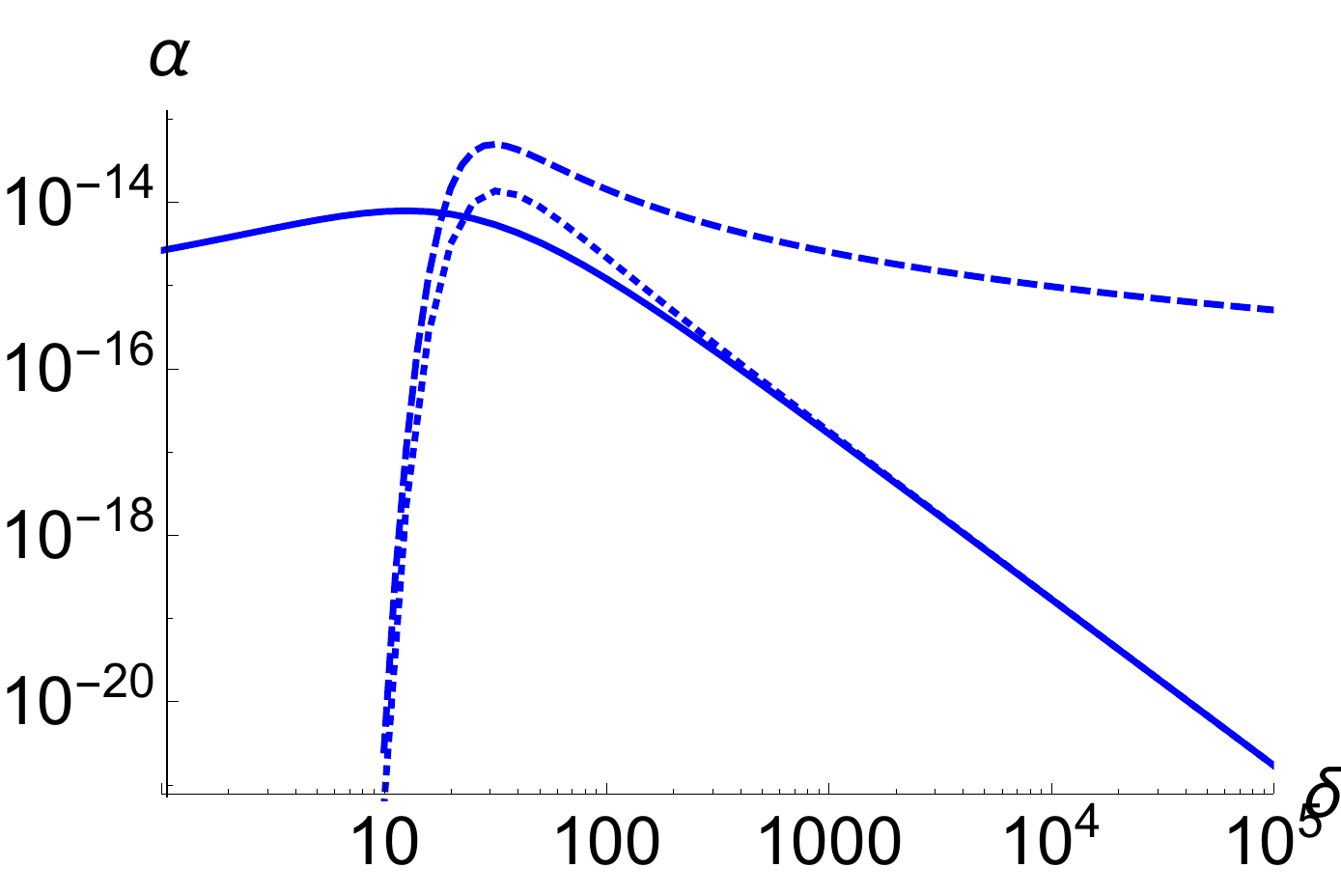}}%
\caption{$r$ vs. $n_s$ (a), $r$ vs. $\delta$ (b), $\delta$ vs. $n_s$ (c), and $\alpha$ vs. $\delta$ (d) with $N_e = 60$ $e$-folds for the CW potential \eqref{eq:U:final}. Blue dashed line represents inflation in region 1), blue dotted line in region 2) and blue continuous in region 3). For reference we plot the lines for quartic (brown), quadratic (black) and $R^2$ (orange) inflation for $N_e \in [50,60]$. The gray areas represent the 1,2$\sigma$ allowed regions from Planck 2018 data~\cite{Planck2018:inflation}.}
\label{fig:CW:60}
\begin{center}
    \includegraphics[width=0.5\textwidth]{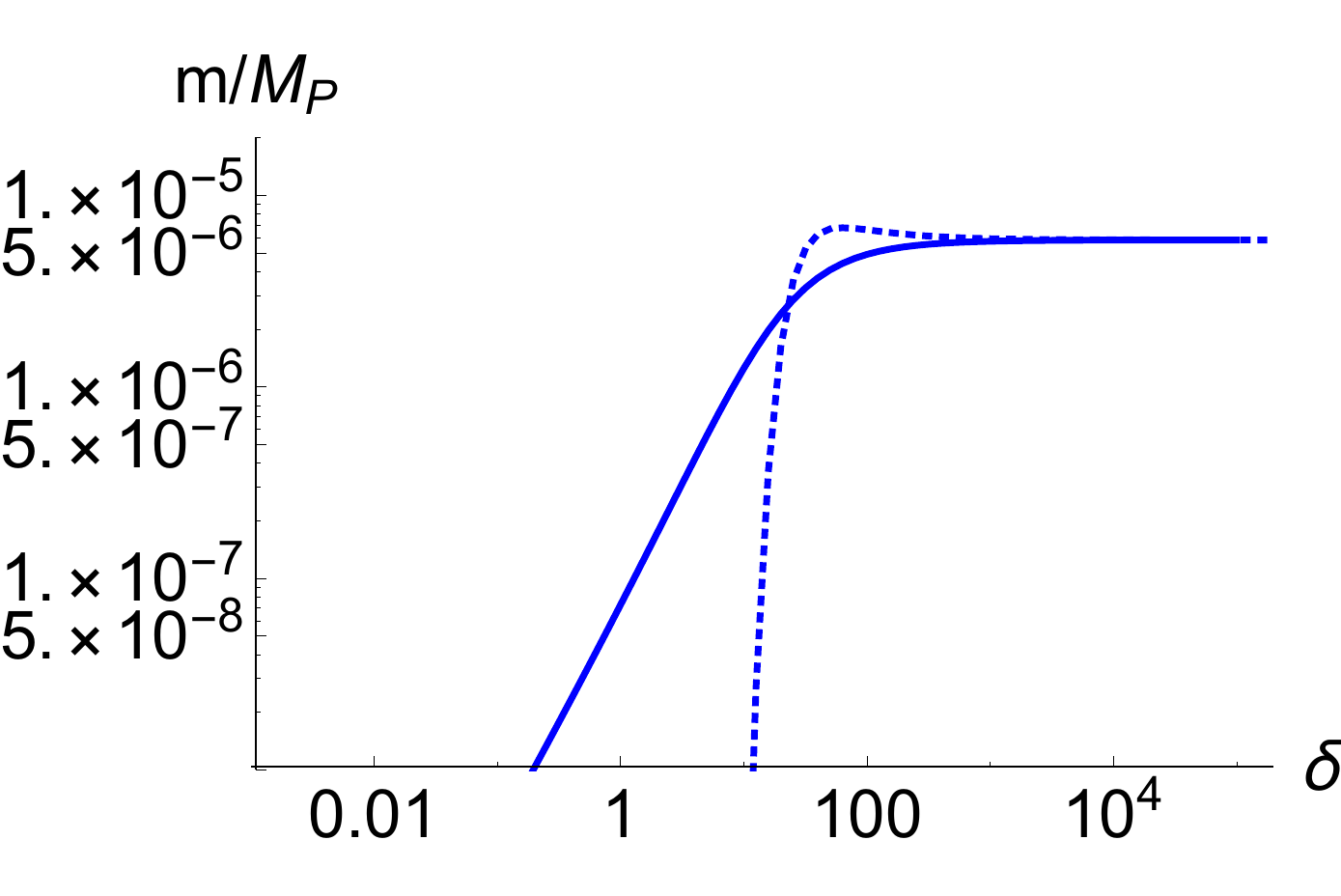}
\end{center}
\caption{Inflaton mass $m$ vs. $\delta$ with $N_e = 60$ $e$-folds for the CW potential \eqref{eq:U:final}. Blue dotted line represents inflation in region 2) and blue continuous in region 3).}
\label{fig:m:CW:60}
\end{figure}
The field value $\chi_N$ at the time a given scale left the horizon is given by the corresponding $N_e$. 
Other two relevant observables, i.e. the spectral index and the tensor-to-scalar ratio are respectively expressed in terms of the slow-roll parameters by
\bea
n_s &\simeq& 1+2\eta-6\epsilon \label{eq:ns} \\
r &\simeq& 16\epsilon \, \label{eq:r} .
\eea
To reproduce the correct amplitude for the curvature power spectrum, the potential has to satisfy \cite{Planck2018:inflation}
\beq
\label{eq:As:constraint}
\ln \left(10^{10} A_s \right) = 3.044 \pm 0.014   \, ,
\ee
where
\beq
 A_s = \frac{1}{24 \pi^2 \MP^4} \frac{U(\chi_N)}{\epsilon(\chi_N)} \label{eq:As} \, .
\ee
This constraint is commonly used to fix the normalization of the inflaton potential.
We can easily verify that the scalar potential in eq. \eqref{eq:U:final} exhibits three stationary points
\begin{equation}
      \chi_{1} 	= 0\,, \qquad     \chi_{2} = M/\sqrt{e}\,, \qquad     \chi_{3} = M\,,
      \label{eq:min:mc}
\end{equation}
where $e$ is Euler's number. As expected, $\chi_2$ is a local maximum while $\chi_{1,3}$ are the degenerate minima required by the MPCP with $U(\chi_{1,3})=0$. As we can see from Fig. \ref{fig:V_and_eps_CW}, there are three possible regions suitable for inflation: 
\begin{itemize}
 \item[1)] backward hilltop inflation, $0 < \chi < \chi_{2} $, where the inflaton slow-rolls from the local maximum back to smaller field values
 \item[2)] forward hilltop inflation, $\chi_{2} < \chi < \chi_{3} $, where the inflaton slow-rolls from the local maximum down to larger field values
 \item[3)] large field inflation, $\chi_{3} < \chi $, the inflaton slow-rolls from ``infinity'' downwards to the second minimum.
\end{itemize}
%
Since we do not know the value of the scale $M$, for convenience we parametrize it as
\begin{equation}
    M = \delta \, M_P \, ,
    \label{eq:M}
\end{equation}
and scan over multiple values of $\delta$. Our results are presented in Figs. \ref{fig:CW:50}-\ref{fig:m:CW:60}. In Fig. \ref{fig:CW:50} we plot $r$ vs. $n_s$ (a), $r$ vs. $\delta$ (b), $\delta$ vs. $n_s$ (c), and $\alpha$ vs. $\delta$ (d) with $N_e = 50$ $e$-folds for the CW potential in eq. \eqref{eq:U:final}. 
Blue dashed line represents inflation in region 1), blue dotted line in region 2) and blue continuous in region 3).
For reference we also plot predictions of Starobinsky (orange), quartic (brown) and quadratic (black) inflation for $N_e \in [50,60]$. The gray areas represent the 1,2$\sigma$ allowed regions coming from Planck 2018 data~\cite{Planck2018:inflation}. The same is for Fig. \ref{fig:CW:60} but for $N_e = 60$ $e$-folds. Even though this second set of results is shifted with the respect to the $N_e = 50$ set, their behaviour is the same. The dashed/continuous line approaches the value of quartic inflation $\alpha \chi^4$ for very high/small $\delta$. This can be explained by looking at the different components of the CW potential in eq. \eqref{eq:U:final}: $\ln^2\left(\chi/M\right)$ and $\chi^4$. When comparing them at $\chi_{N}$ and $\chi_{f}$ with high/small $\delta$ values we see that
\begin{gather}
    \frac{\ln^2\left(\frac{\chi_{f}}{M}\right)}{\ln^2\left(\frac{\chi_{N}}{M}\right)}\approx 1\,,\\
    \left(\frac{\chi_{f}}{\chi_{N}}\right)^4 << 1\,.
\end{gather}
Therefore, the contribution of the logarithmic component is subdominant compared to the one of the $\chi^4$ part, meaning that, for such values of $\delta$, the potential is essentially quartic during inflation. Also at very high $\delta$ values the continuous and dotted lines approach quadratic inflation. In this case inflation is happening very close to second minimum, therefore it is expected that the potential behaves quadratically around the minimum $\chi_3 = M$.

As mentioned before, the behaviour of the results for $N_e=50$ and $N_e=60$ is the same. However only for $N_e=60$ some results are in the allowed region. This happens for inflation taking place in region 2). From Fig. \ref{fig:CW:60} we can see that this happen when $40 \lesssim \delta \lesssim 70$ i.e. $M$ is trans-Planckian. First of all we stress that such a trans-Planckian scale is just an artifact of the chosen parametrization. The full argument of a logarithmic term generated by a radiative correction is of the form $g \chi/\mu$ where $\mu$ is an energy scale and $g$ some coupling constant. Therefore it is straightforward to check that, when the theory is perturbative ($g<1$), by identifying $M=\mu/g$, $M$ can be easily trans-Planckian even though $\mu$ is not, provided a small enough $g$. However, since we have the appearance of an effective trans-Planckian scale, it is worth to check that the inflaton mass remains sub-Planckian. Such a mass is defined as the second derivative of the potential evaluated at the minimum
\beq
 m^2 = \left. \frac{\partial^2 U}{\partial \chi^2} \right|_{\chi=\chi_{\rm min}}
\eeq
Since we have 2 degenerate minima, $\chi_1=0$ and $\chi_3=M$, the inflaton mass changes according to where inflation happens. For region 1) we have that $m=0$, while for region 2) and 3) we obtain $m= \sqrt{2 \alpha} \, M$. The results for the inflaton mass for region 2) and 3) for $N_e=50$ and $N_e=60$ are given respectively in Figs. \ref{fig:m:CW:50} and \ref{fig:m:CW:60}. We can see that the inflaton mass is always sub-Planckian, ensuring the consistency of the model.

However, as mentioned before, the allowed region of this model is relatively small. On the other hand it is well known that non-minimal couplings between inflation and gravity arise via radiative corrections. Such couplings may strongly affect all the predictions. This will be studied in the next section.

\section{Non-minimal CW inflation and multiple-point criticality principle } \label{sec:xiCW}
We consider the following Jordan frame action
\begin{equation}
S = \!\! \int \!\! d^4x \sqrt{-g^J}\left(-\frac{M_P^2}{2}f(\phi)R(\Gamma) + \frac{(\partial \phi)^2}{2}  - V(\phi) \right) ,
\label{eq:JframeL}
\end{equation}
which is essentially the action given in eq. \eqref{eq:L:classic} where we made explicit the dependence of the Ricci scalar $R$ from the connection $\Gamma$ and we added the non-minimal coupling to gravity $f(\phi)$
\begin{equation}
f(\phi)=1 + \xi \frac{\phi^2}{M_P^2} \, ,  
  \label{eq:f:H}
\end{equation}
which is the usual Higgs-inflation \cite{Bezrukov:2007ep} non-minimal coupling but not necessarily identifying the inflaton with the Higgs boson. 
We also relabelled the inflaton as $\phi$ and its potential as $V$ so that now the effective potential\footnote{While cosmological perturbations are invariant under frame transformations (see for instance \cite{Prokopec:2013zya,Jarv:2016sow}), the equivalence of the Einstein and Jordan frames at the quantum level is still to be established. In the present article we therefore apply the following strategy: first we compute the effective potential in the Jordan frame, eq.~\eqref{eq:V:final}, and consequently we move to the Einstein frame for computing the slow-roll parameters. Given a scalar potential in the Jordan frame, the cosmological perturbations are then independent, in the slow-roll approximation, of the choice of the frame in which the inflationary observables are evaluated~\cite{Prokopec:2013zya,Jarv:2016sow}. For further discussions on frames equivalence and/or loop corrections in scalar-tensor theories we refer the reader to Refs.~\cite{Jarv:2014hma,Kuusk:2015dda,Kuusk:2016rso,Flanagan:2004bz,Catena:2006bd,Barvinsky:2008ia,DeSimone:2008ei,Barvinsky:2009fy,Barvinsky:2009ii,Steinwachs:2011zs,Chiba:2013mha,George:2013iia,Postma:2014vaa,
Kamenshchik:2014waa,George:2015nza,Miao:2015oba,Buchbinder:1992rb,Elizalde:1993ee,Elizalde:1993ew,Elizalde:1994im,Inagaki:2015fva,Burns:2016ric,Fumagalli:2016lls,Artymowski:2016dlz,Fumagalli:2016sof,Bezrukov:2017dyv,Karam:2017zno,Narain:2017mtu,Ruf:2017xon,Markkanen:2017tun,Markkanen:2018bfx,Ohta:2017trn,Ferreira:2018itt,Karam:2018squ}.}
has the same functional form as before
\beq
    V(\phi) = \alpha \ln^2\! \left(\frac{\phi }{M}\right) \phi ^4 \, ,
    \label{eq:V:final}
\eeq
but in the Jordan frame. In order to keep the notation consistent with the previous section, from now on $g^{J}_{\mu \nu}$, $\phi$, $V$ will be respectively the metric tensor, the canonically normalized inflaton and its scalar potential in the Jordan frame, while $g_{\mu \nu}$, $\chi$, $U$ are the corresponding counterparts in the Einstein frame. Such a frame is obtained via the Weyl transformation
\begin{eqnarray}
\label{eq:gE}
g_{\mu \nu} = f(\phi) \ g^J_{\mu \nu} \, .
\end{eqnarray}
and it is exactly described by the action given in eq. \eqref{eq:L:classic} where the Einstein frame scalar potential is given by
\beq
U(\chi) = \frac{V(\phi(\chi))}{f^{2}(\phi(\chi))} \, 
\label{eq:U:general}
\eeq
The corresponding canonically normalized field depends on the function $f(\phi)$ and on the gravity formulation under consideration.
In the usual metric case we have
\begin{equation}
\frac{\partial \chi}{\partial \phi} = \sqrt{\frac{3}{2}\left(\frac{M_P}{f}\frac{\partial f}{\partial \phi}\right)^2+\frac{1}{f}} = \frac{M_P \sqrt{M_P^2+\xi  (1+6 \xi) \phi ^2}}{M_P^2+\xi  \phi ^2} \, ,  
  \label{eq:dphim}
\end{equation}
where the first term comes from the transformation of the Jordan frame Ricci scalar and the second from the rescaling of the Jordan frame scalar field kinetic term. 
On the other hand, in the Palatini case \cite{Bauer:2008zj}, the field redefinition is induced only by the rescaling of the inflaton kinetic term i.e.
\begin{equation}
\frac{\partial \chi}{\partial \phi} = \sqrt{\frac{1}{f}} = \frac{M_P}{\sqrt{M_P^2 + \xi \phi^2}}\, ,  
  \label{eq:dphiP}
\end{equation}
where there is no contribution from the Jordan frame Ricci scalar. Unfortunately it is not always possible to obtain exactly\footnote{This is particularly true for the metric formulation. On the other hand, in the Palatini case, it is also possible to solve and invert exactly the field redefinition obtaining
\begin{equation}
   \phi = \frac{M_P \: \text{sinh} \left (\frac{\sqrt{\xi} \chi}{M_P} \right)}{\sqrt{\xi}} \, ,
    \label{eq:phi:chi:P}
\end{equation}
where we used the boundary condition $\chi(0)=0$. In this case the Einstein frame potential becomes
\begin{equation}
    U(\chi) = \frac{M_P^4 \: \alpha \: \ln^2 \left(\frac{\text{sinh} \left(\frac{\sqrt{\xi} \chi}{M_P}\right)}{\sqrt{\xi} \delta}\right) \: \text{tanh}^4 \left(\frac{\sqrt{\xi} \chi}{M_P}\right)}{\xi^2} \, ,
    \label{eq:U:P}
\end{equation}
where we have used eq. \eqref{eq:M}. } $\phi(\chi)$, however all the phenomenological parameters given in eqs. \eqref{eq:Ne}, \eqref{eq:ns}, \eqref{eq:r} and \eqref{eq:As} can derived using $\phi$ as computational variable, the chain rule $\frac{\partial}{\partial \chi} = \frac{\partial}{\partial \phi} \frac{\partial \phi}{\partial \chi}$ and eq. \eqref{eq:dphim} or \eqref{eq:dphiP}. 

Regardless of the Jordan frame formulation (metric or Palatini) the Einstein frame potential exhibits again three stationary points. The exact canonically normalized value of such points depends on whether we are dealing with metric or Palatini gravity in the Jordan frame. However, since the difference between the two frames relies all in the field redefinition (either \eqref{eq:dphim} or \eqref{eq:dphiP}), using $\phi$ as a computational variable, the position of the stationary points remains the same in both metric and Palatini gravity. By solving $\partial U/\partial \phi =0$, we obtain the following three extremes
\begin{equation}
      \phi_{1} 	= 0\,, \qquad     \phi_{2} = \frac{M_P}{\sqrt{\xi }} \sqrt{W\left(\frac{ \xi }{e}\delta^2\right)}\,, \qquad     \phi_{3} = M=\delta M_P\,,
\label{eq:min:nmc}
\end{equation}
where $W(z)$ gives the principal solution for $w$ in $z=w e^w$. Since the minima correspond to $U(\phi)=V(\phi)/f^2(\phi)=0$, their position is unchanged with respect to minimally coupled case in eq. \eqref{eq:min:mc}. On the other hand, the position of the local maximum is changed because of the contribution of $\partial f/\partial \phi$ in  $\partial U/\partial \phi$. As before, the general behaviour of $U$ is still depicted by Fig. \ref{fig:V_and_eps_CW} and analogously we can identify three possible regions for inflation
\begin{itemize}
 \item[1)] backward hilltop inflation, $0 < \phi < \phi_{2} $
 \item[2)] forward hilltop inflation, $\phi_{2} < \phi < \phi_{3} $
 \item[3)] large field inflation, $\phi_{3} < \phi $
\end{itemize}
which we describe separately in the following subsections.

\subsection{Backward hilltop inflation: $0 < \phi < \phi_{2} $}
\begin{figure}[p]
    \begin{center}
    \vspace{-0.75cm}
    \subfloat[]{\includegraphics[width=0.44\textwidth]{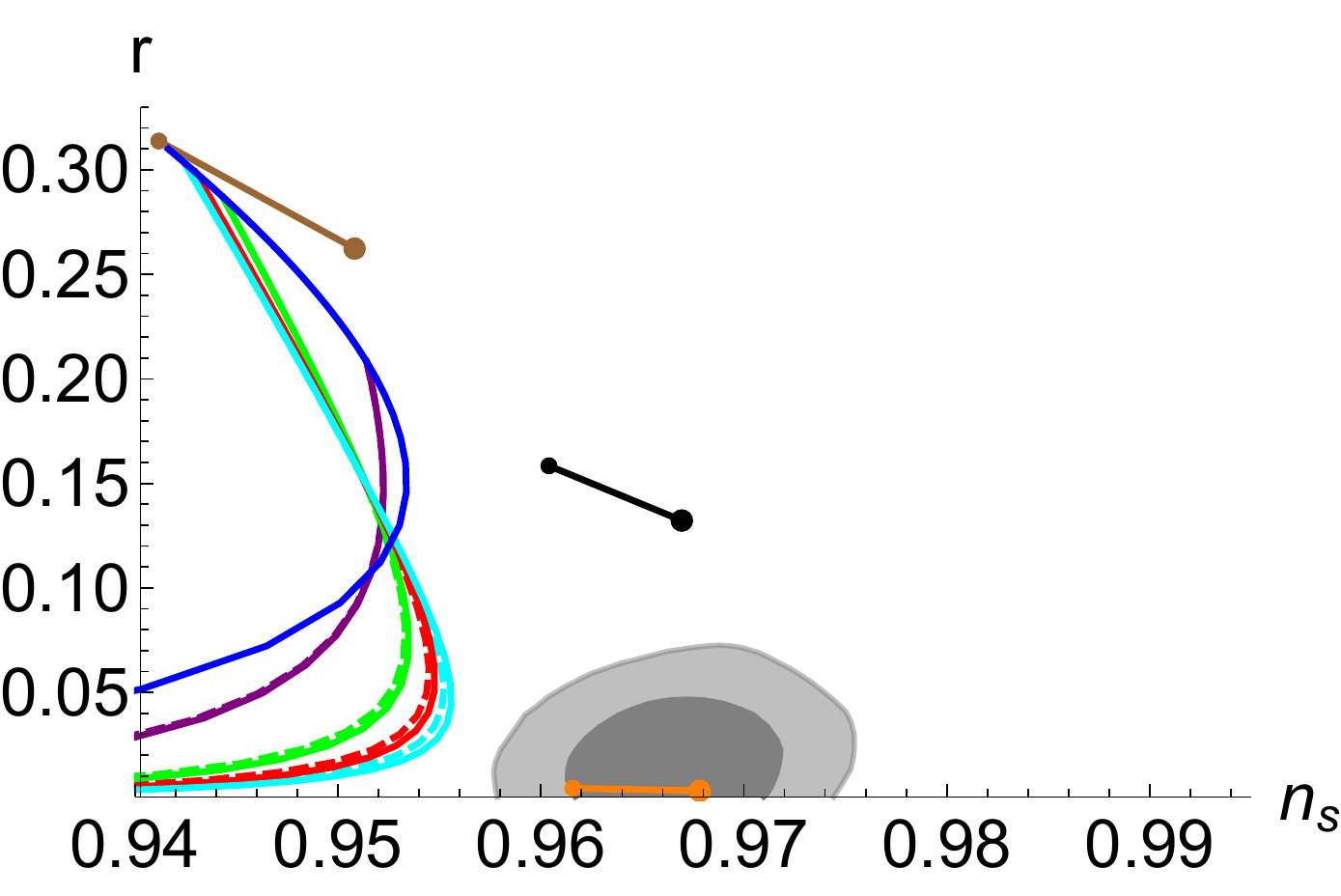}}%
    \subfloat[]{\includegraphics[width=0.44\textwidth]{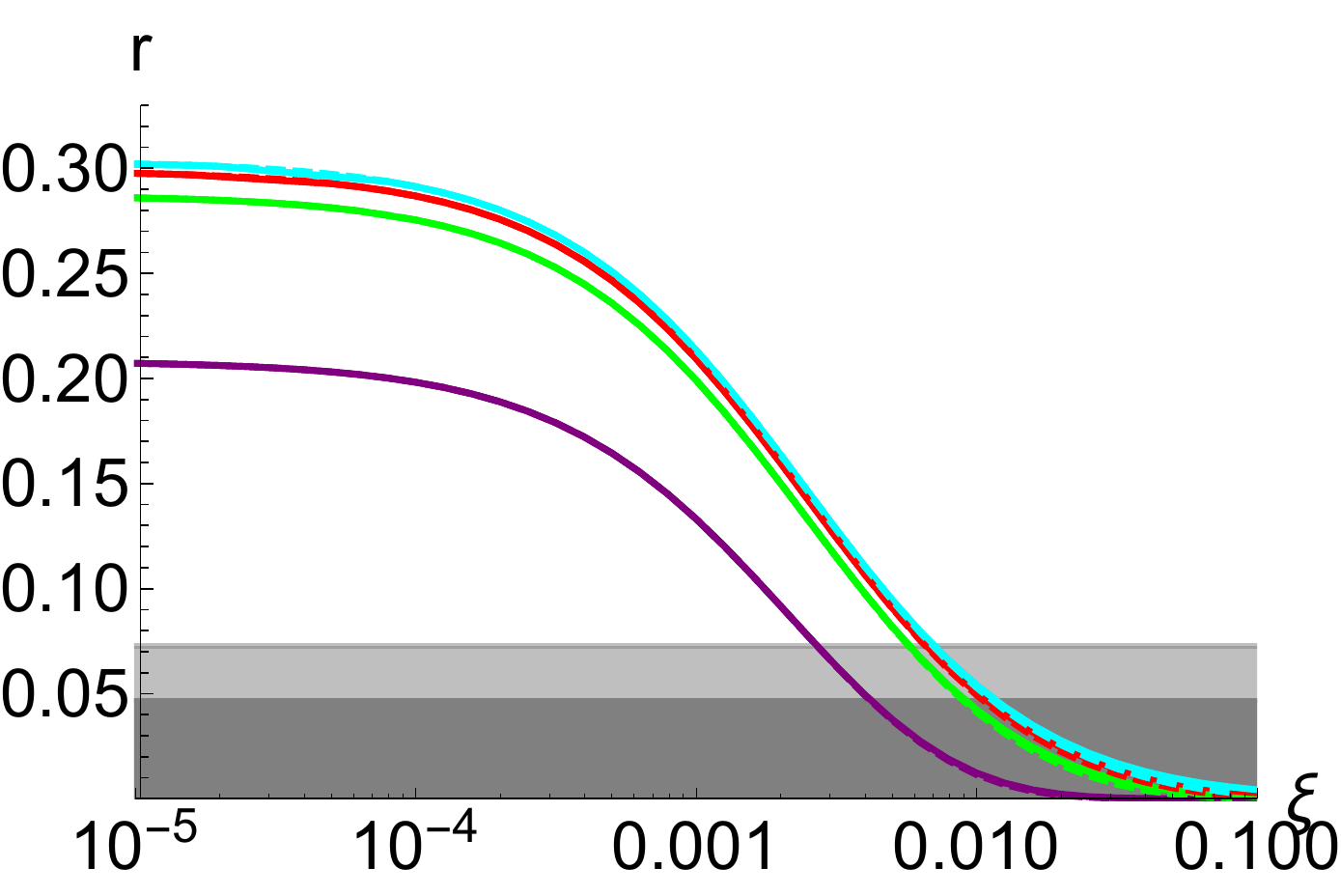}}%
   \\
   \vspace{-0.75cm}
    \subfloat[]{\includegraphics[width=0.44\textwidth]{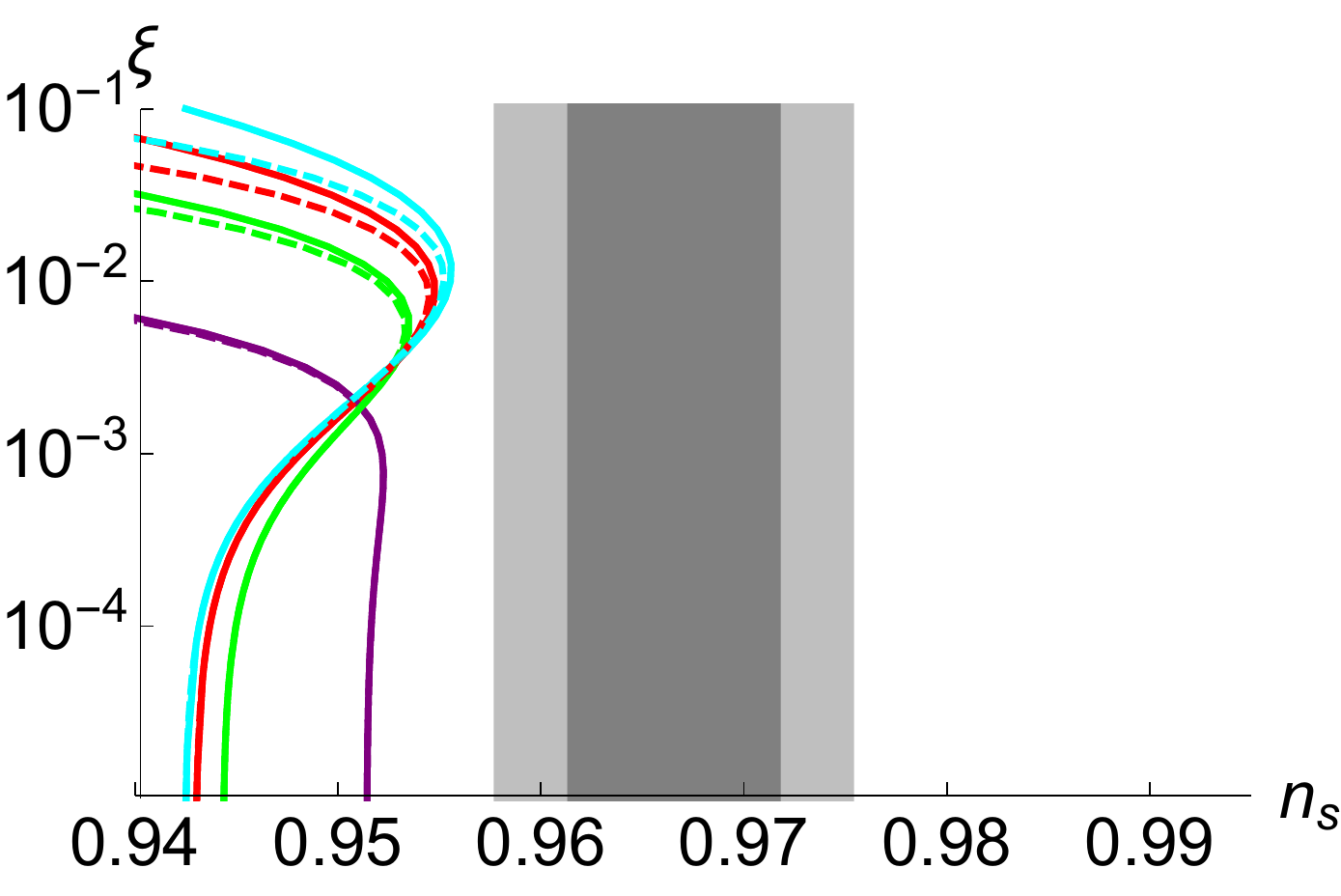}}%
    \subfloat[]{\includegraphics[width=0.44\textwidth]{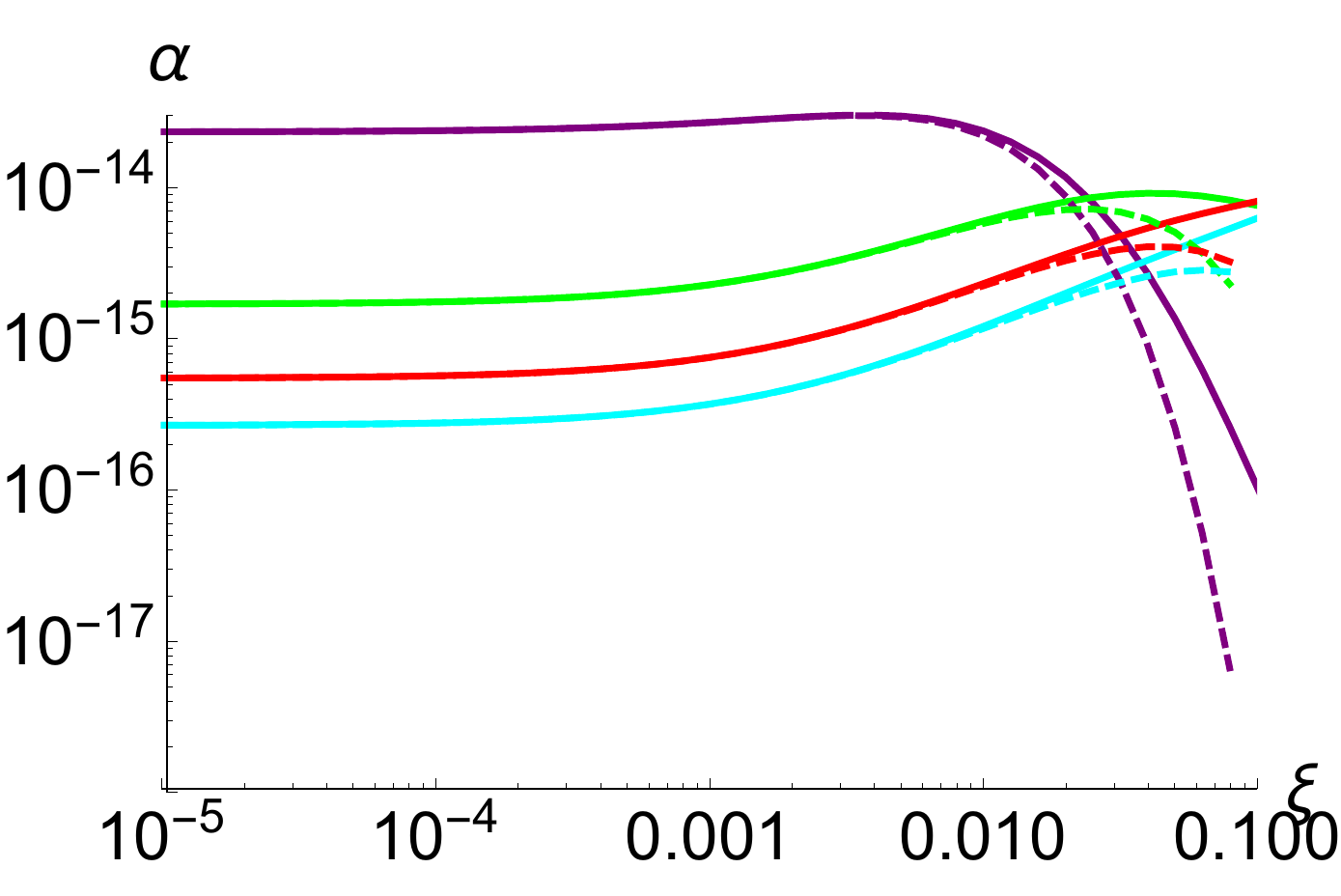}}\\
    \vspace{-0.5cm}
    \subfloat[]{\includegraphics[width=0.44\textwidth]{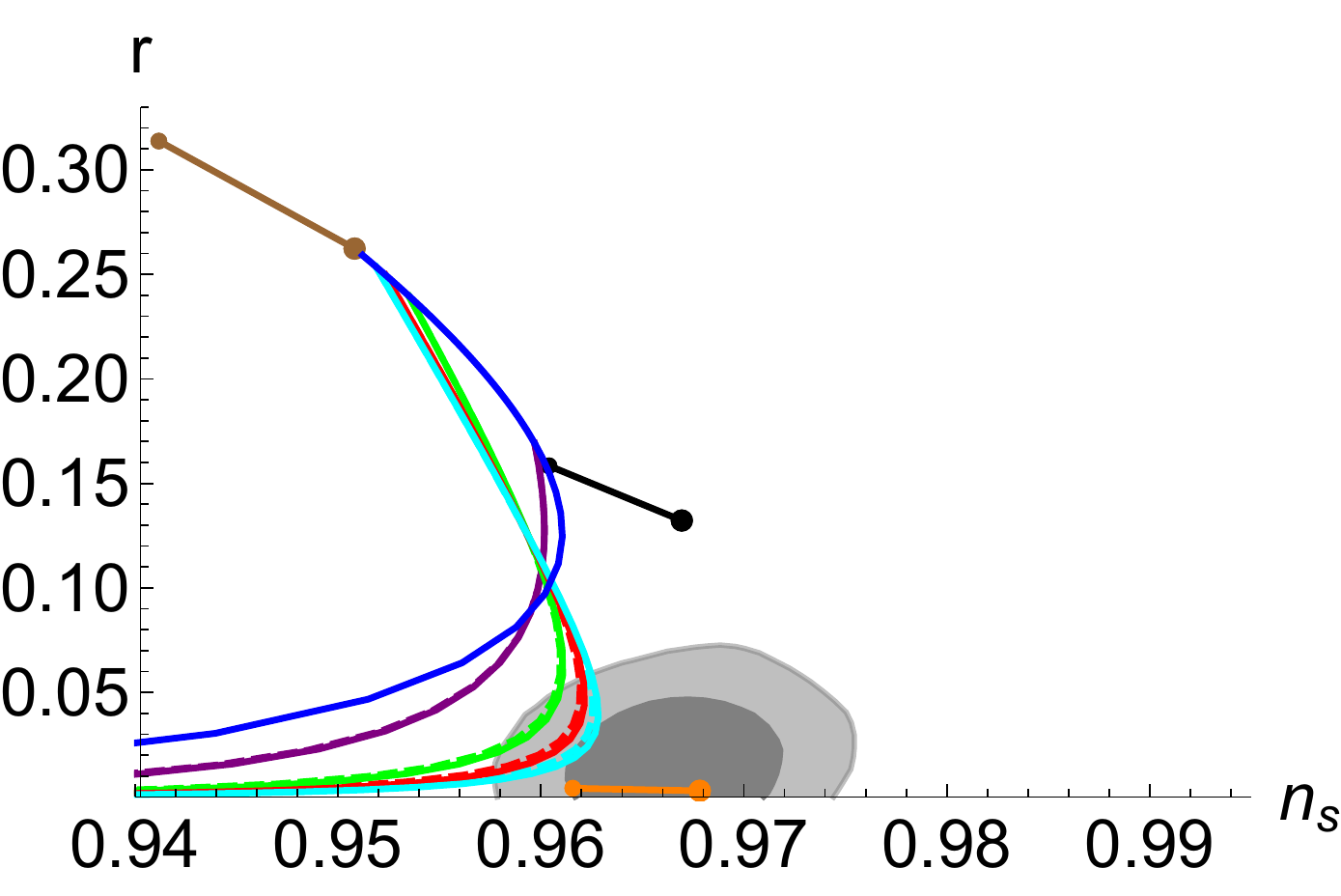}}%
    \subfloat[]{\includegraphics[width=0.44\textwidth]{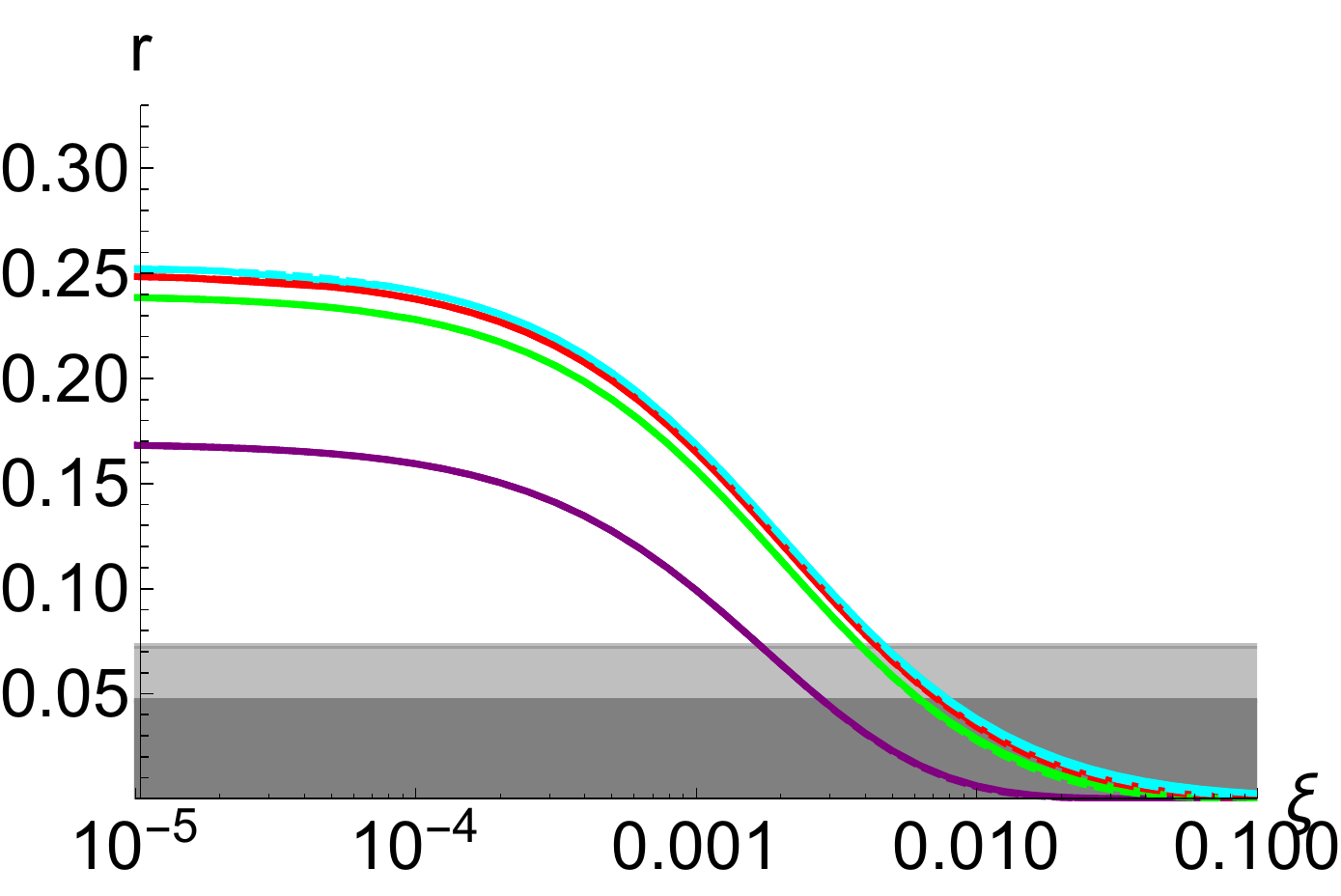}}%
    \\
    \vspace{-0.75cm}
    \subfloat[]{\includegraphics[width=0.44\textwidth]{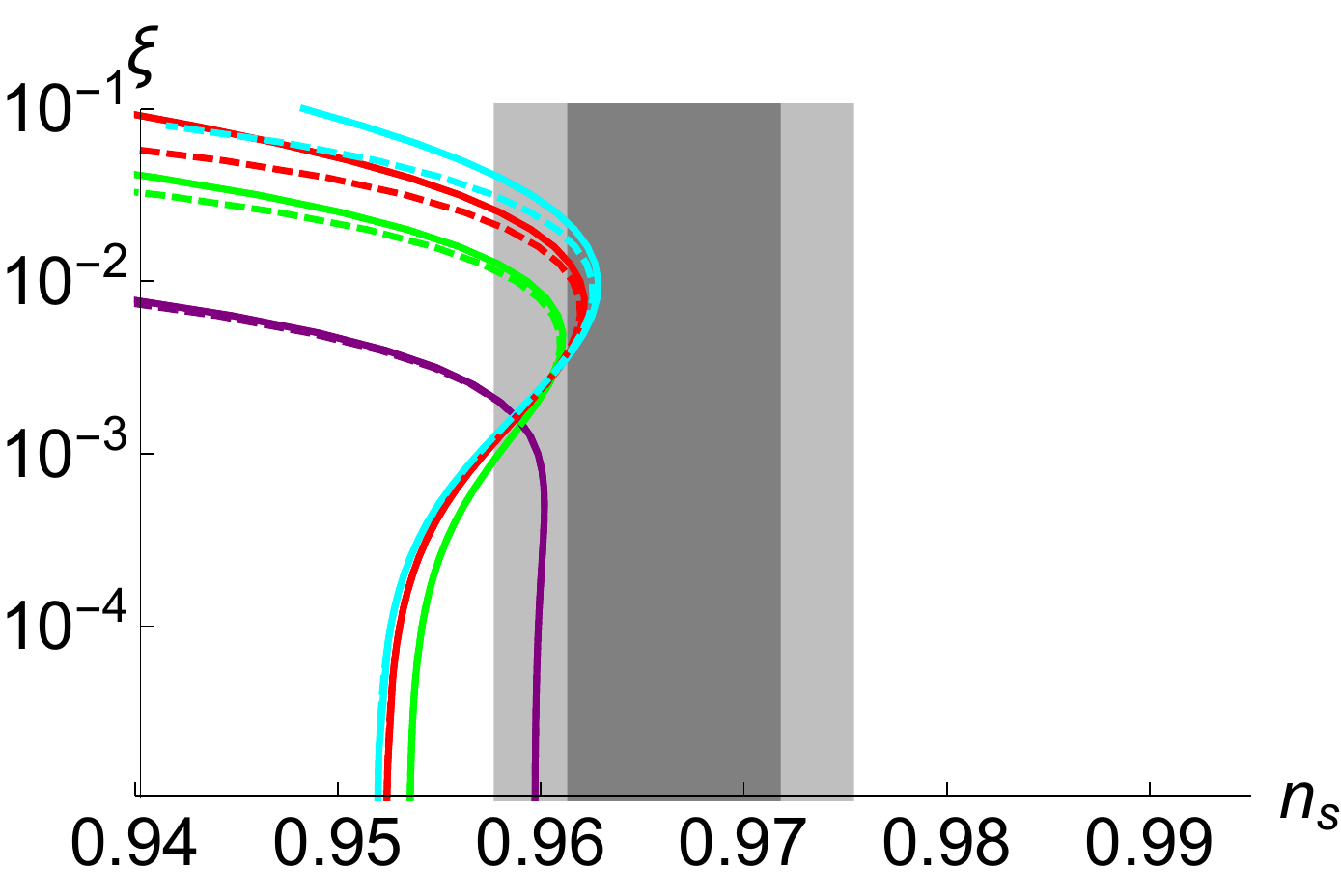}}%
    \subfloat[]{\includegraphics[width=0.44\textwidth]{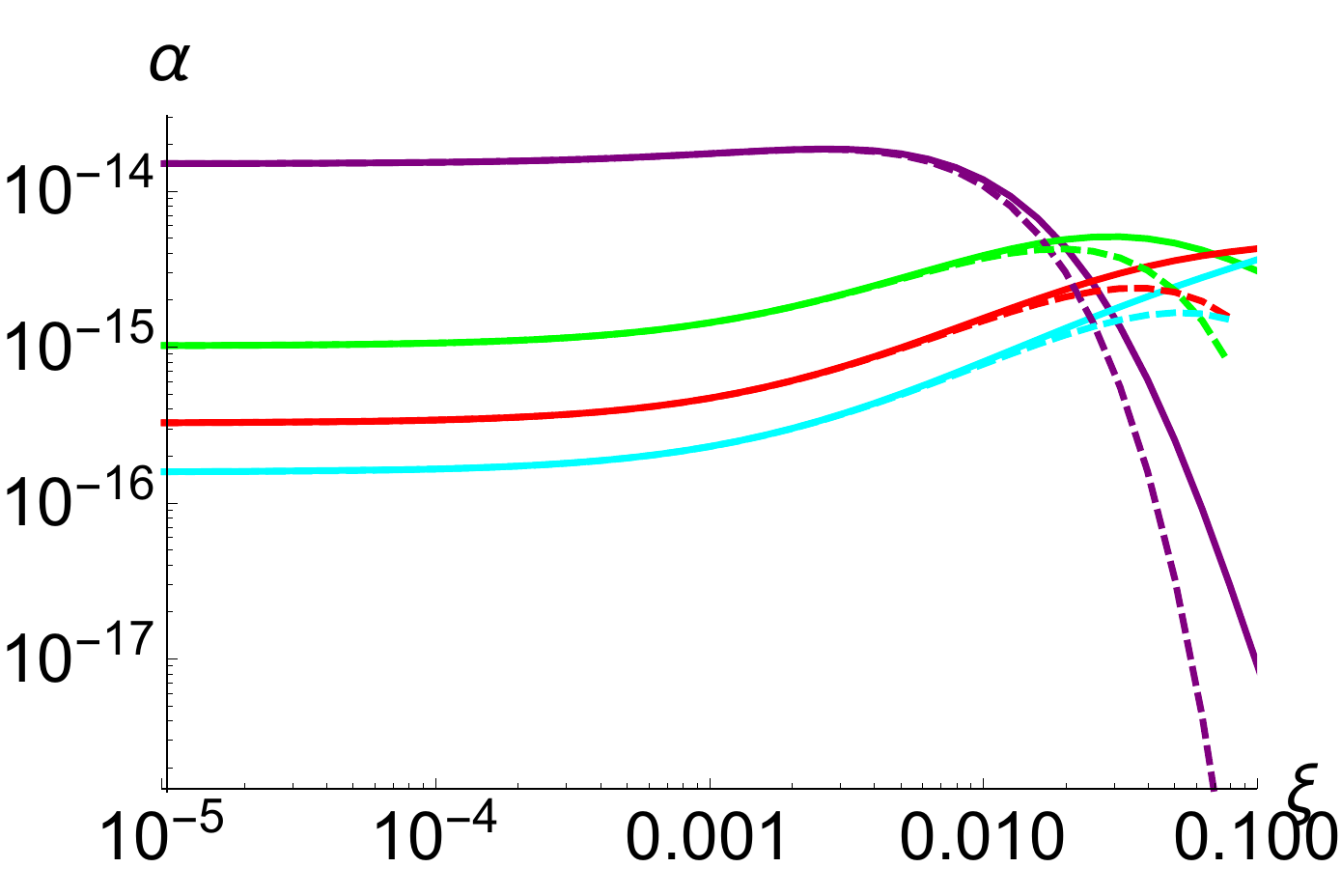}}%
    \end{center}
    \vspace{-0.6cm}
   \caption{$r$ vs. $n_s$ (a,e), $r$ vs. $\xi$ (b,f), $\xi$ vs. $n_s$ (c,g) and $\alpha$ vs. $\xi$ (d,h) for $\delta=10^{2,4,6,8}$ respectively in purple, green, red and cyan with $N_e =50,60$ $e$-folds in region 1. Continuous/dashed line represents metric/Palatini gravity. For reference we plot the predictions of CW inflation for $N_e=50,60$ (blue), $R^2$ (orange), $\chi^2$ (black) and $\chi^4$ (brown) inflation for $N_e \in [50,60]$. The gray areas represent the 1,2$\sigma$ allowed regions from Planck 2018 data~\cite{Planck2018:inflation}.}%
   \label{fig:results:1}
\end{figure}

In this subsection we describe the phenomenology for inflation happening from the local maximum backward to zero, i.e. in the region $0 < \phi < \phi_{2} $. In Fig. \ref{fig:results:1} we show $r$ vs. $n_s$ (a), $r$ vs. $\xi$ (b), $\xi$ vs. $n_s$ (c) and $\alpha$ vs. $\xi$ (d) for the reference values of $\delta=10^2$ (purple),  $\delta=10^4$ (green) , $\delta=10^6$ (red), $\delta=10^8$ (cyan)  with $N_e =50$ $e$-folds. 
Continuous line represents metric gravity, while dashed line stands for Palatini gravity.
For reference we plot the predictions of CW inflation for $N_e=50$ (blue), Starobinsky (orange), quadratic (black) and quartic (brown) inflation for $N_e \in [50,60]$.  The gray areas represent the 1,2$\sigma$ allowed regions coming from Planck 2018 data~\cite{Planck2018:inflation}. Figs.  \ref{fig:results:1}(e),(f),(g),(h) are the same as Figs. \ref{fig:results:1}(a),(b),(c),(d) but for $N_e =60$ $e$-folds. First of all we notice that the $r$ vs. $n_s$ results of metric and Palatini gravity are quite similar. This is because in the plotted region $\xi$ is relatively small ($\xi \lesssim 0.1$), therefore the difference in the field redefinitions \eqref{eq:dphim} and \eqref{eq:dphiP} does not play a relevant role in this specific plot. However it is still possible to appreciate some difference in the $\xi$ vs. $n_s$ and $\alpha$ vs. $\xi$ plots for $\xi \gtrsim 0.01$. At a given $\delta$ the effect of the non-minimal coupling is, as usual, to drive $r$ towards smaller values with $\xi$ increasing, while $n_s$ is first driven towards larger values, reaches a maximum and then moves towards lower values far away from the allowed Planck region. For $\delta \gtrsim 10^4$, the results behave like a regular non-minimally coupled quartic inflation, until the aforementioned effect kicks in and drives away the results towards lower values of $n_s$. Since $M=\delta M_P$ takes trans-Planckian values, we should check the value of the inflaton mass. As before, since in this case the inflaton is slow-rolling towards 0, the inflaton mass is 0 as well. As final remark, we notice that, as usual, the difference in the number of $e$-folds does not affect the general behaviour of the results but only their eventual agreement with the observational constraints. In this case, only $N_e=60$ is allowed when $\delta \gtrsim 10^4$ and $\xi$ is around $0.01$. 

\subsection{Forward hilltop inflation: $\phi_2 < \phi < \phi_3 $}
\begin{figure}[p]%
    \centering
    \subfloat[]{\includegraphics[width=0.46\textwidth]{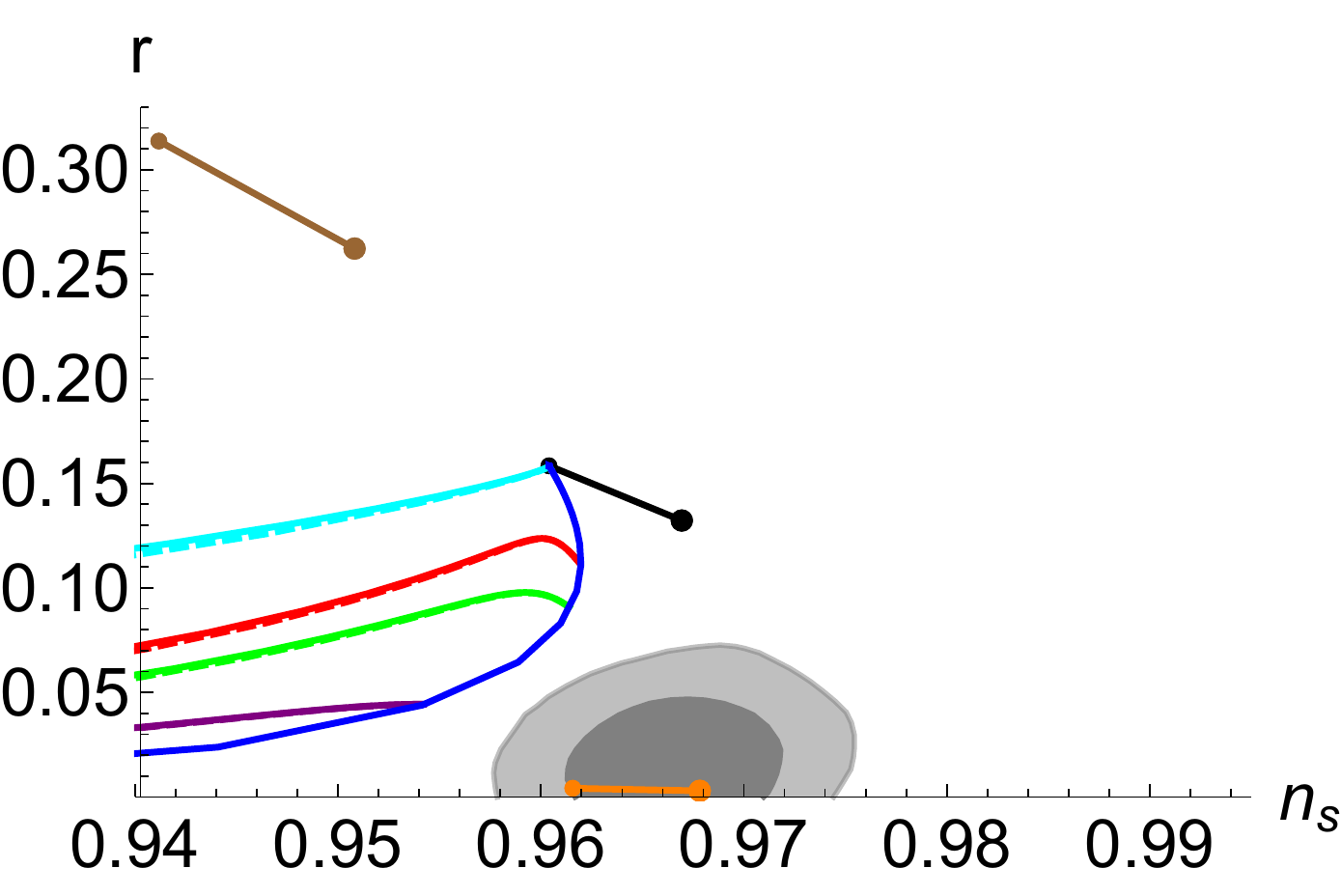}}%
    \subfloat[]{\includegraphics[width=0.46\textwidth]{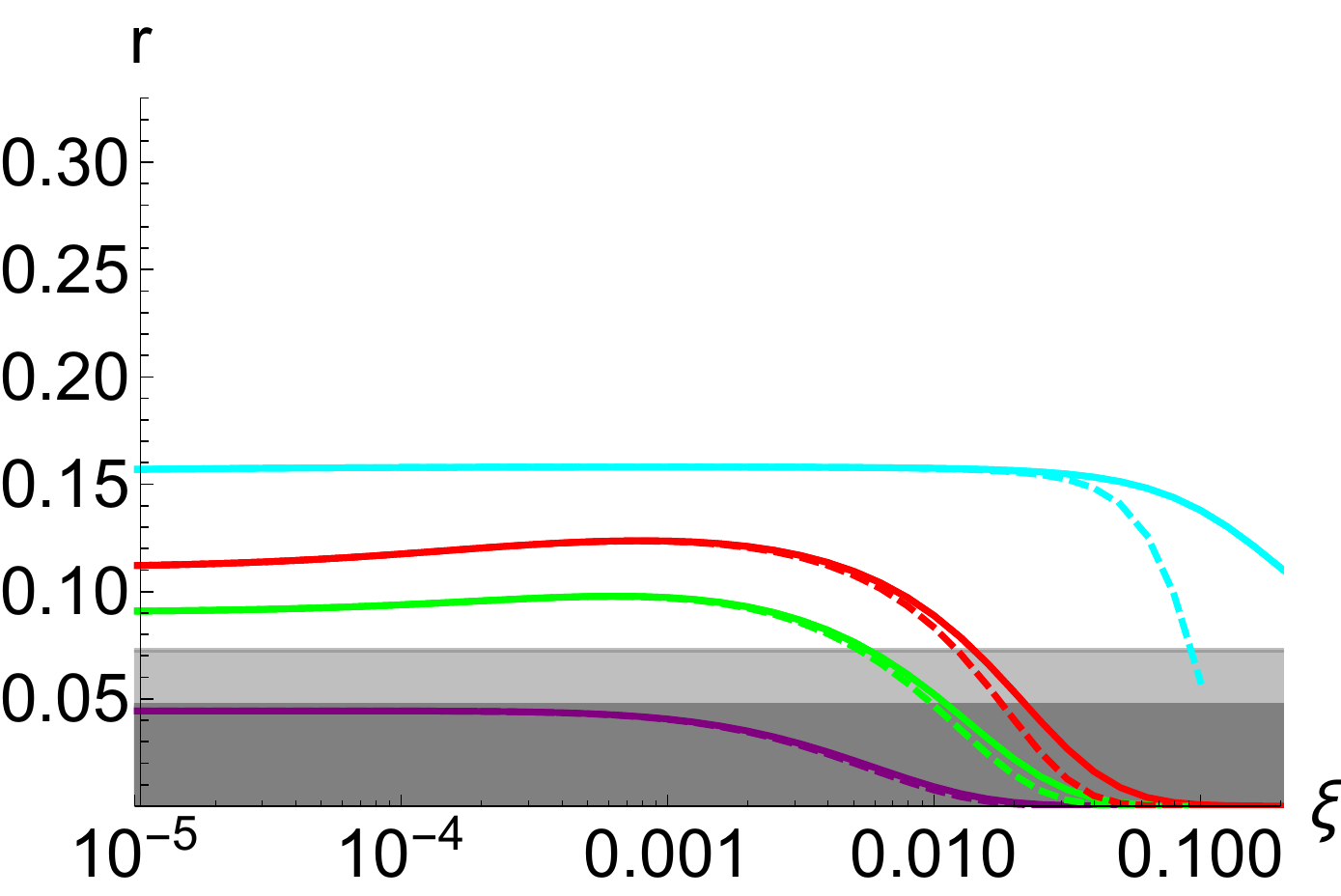}}\\
    \subfloat[]{\includegraphics[width=0.46\textwidth]{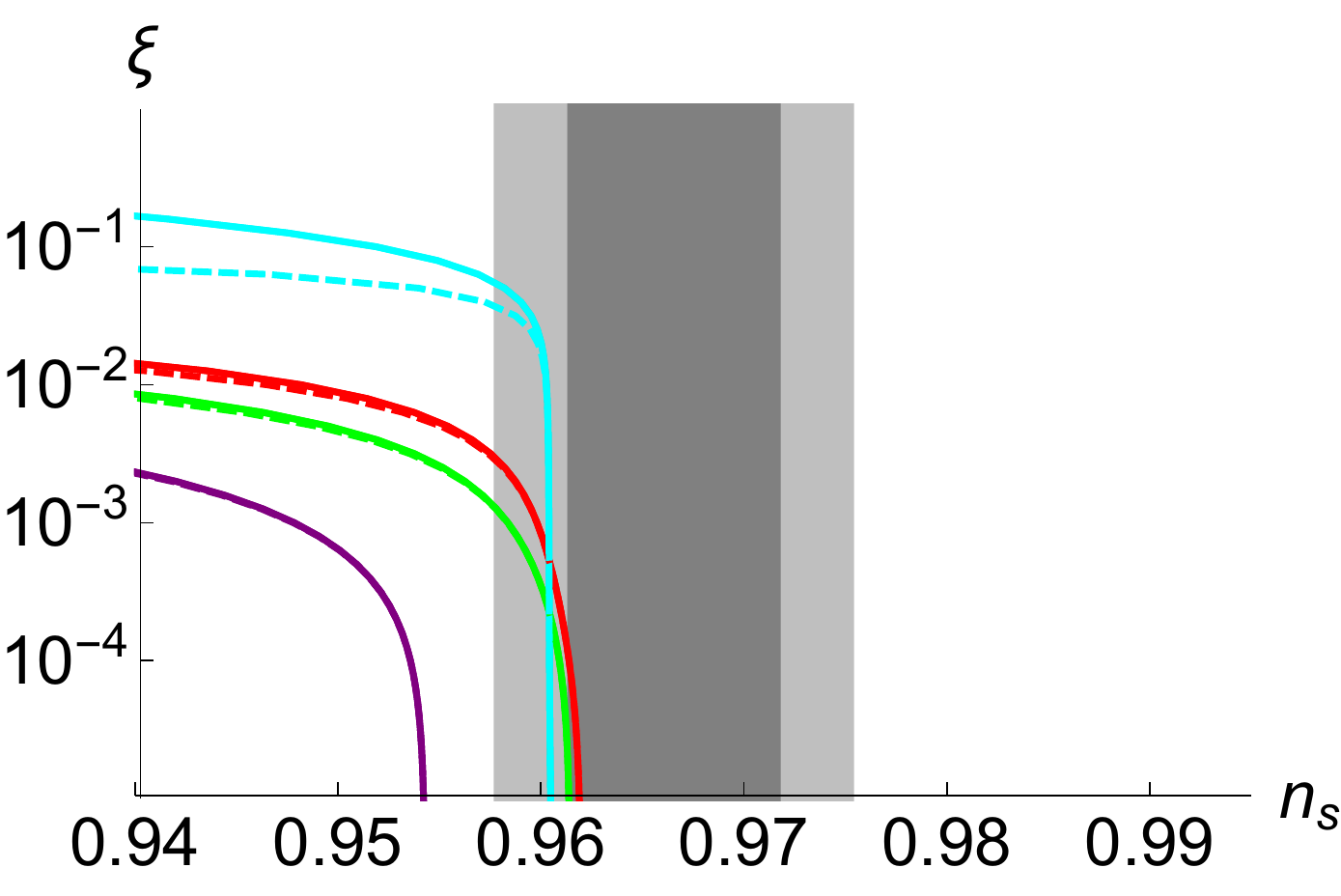}}%
    \subfloat[]{\includegraphics[width=0.46\textwidth]{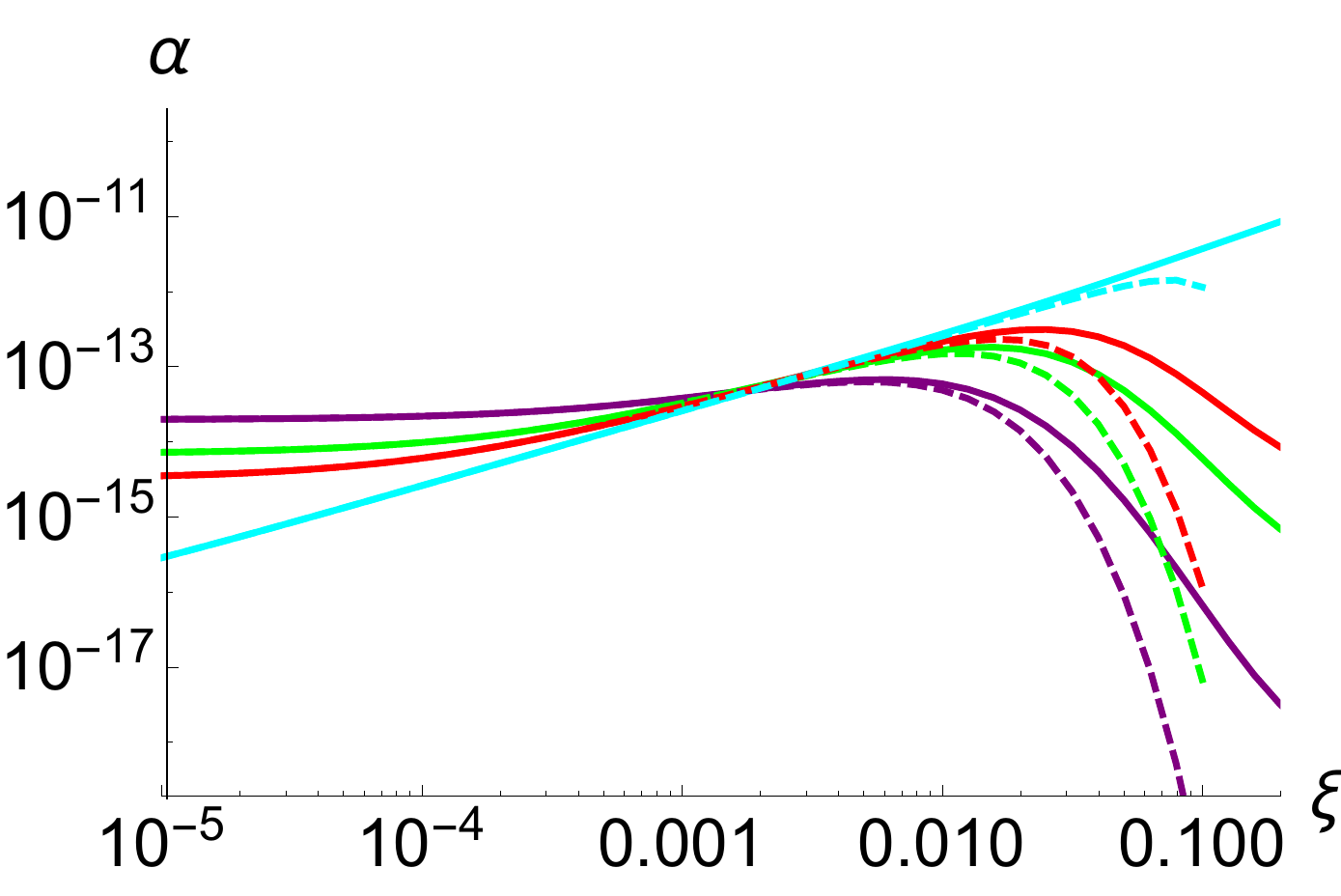}}%
    \caption{$r$ vs. $n_s$ (a), $r$ vs. $\xi$ (b), $\xi$ vs. $n_s$ (c) and $\alpha$ vs. $\xi$ (d) for $\delta=40,70,100,1000$ respectively in purple, green, red and cyan with $N_e =50$ $e$-folds in region 2. Continuous/dashed line represents metric/Palatini gravity. For reference we plot the predictions of CW inflation for $N_e=50$ (blue), $R^2$ (orange), $\chi^2$ (black) and $\chi^4$ (brown) inflation for $N_e \in [50,60]$. The gray areas represent the 1,2$\sigma$ allowed regions from Planck 2018 data~\cite{Planck2018:inflation}.}%
\label{fig:results:2:50}
\vskip 1cm
\begin{center}
    \includegraphics[width=0.46\textwidth]{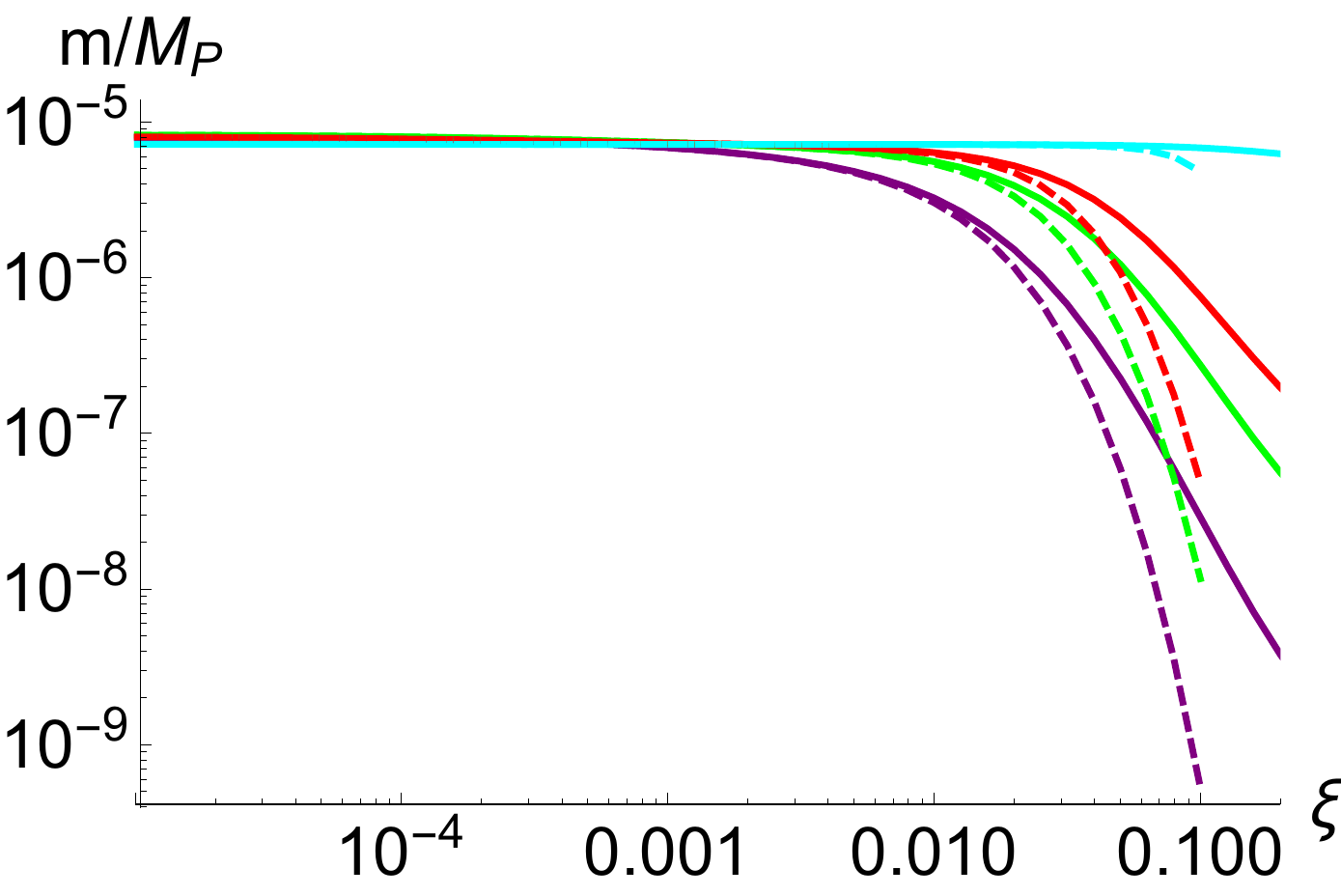}
\end{center}
\caption{$m$ vs. $\xi$ for $\delta=40,70,100,1000$ respectively in purple, green, red and cyan with $N_e =50$ $e$-folds in region 2. Continuous/dashed line represents metric/Palatini gravity.}
\label{fig:mvsxi:2:50}
\end{figure}

\begin{figure}[p]%
    \centering
    \subfloat[]{\includegraphics[width=0.46\textwidth]{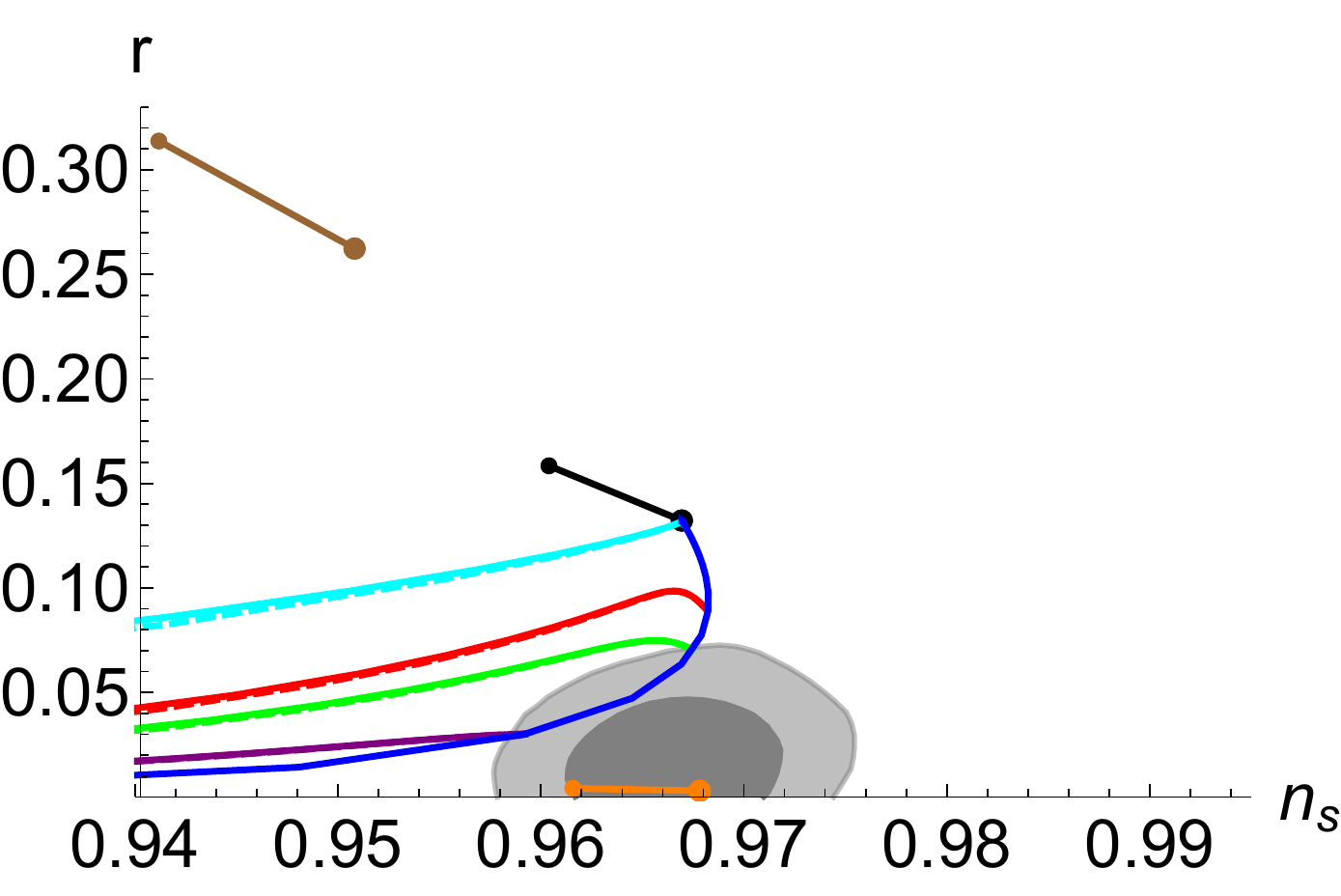}}%
    \subfloat[]{\includegraphics[width=0.46\textwidth]{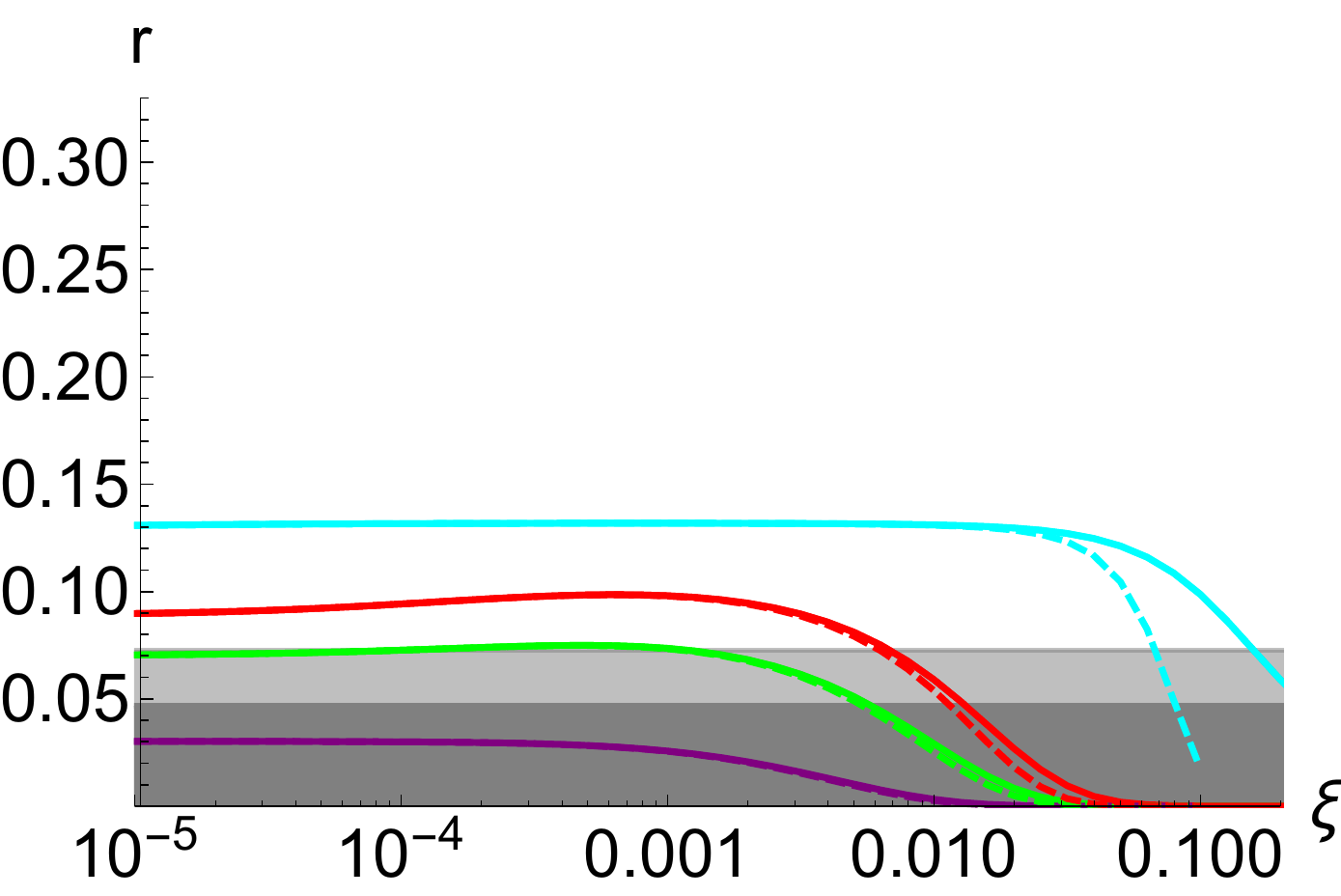}}\\
    \subfloat[]{\includegraphics[width=0.46\textwidth]{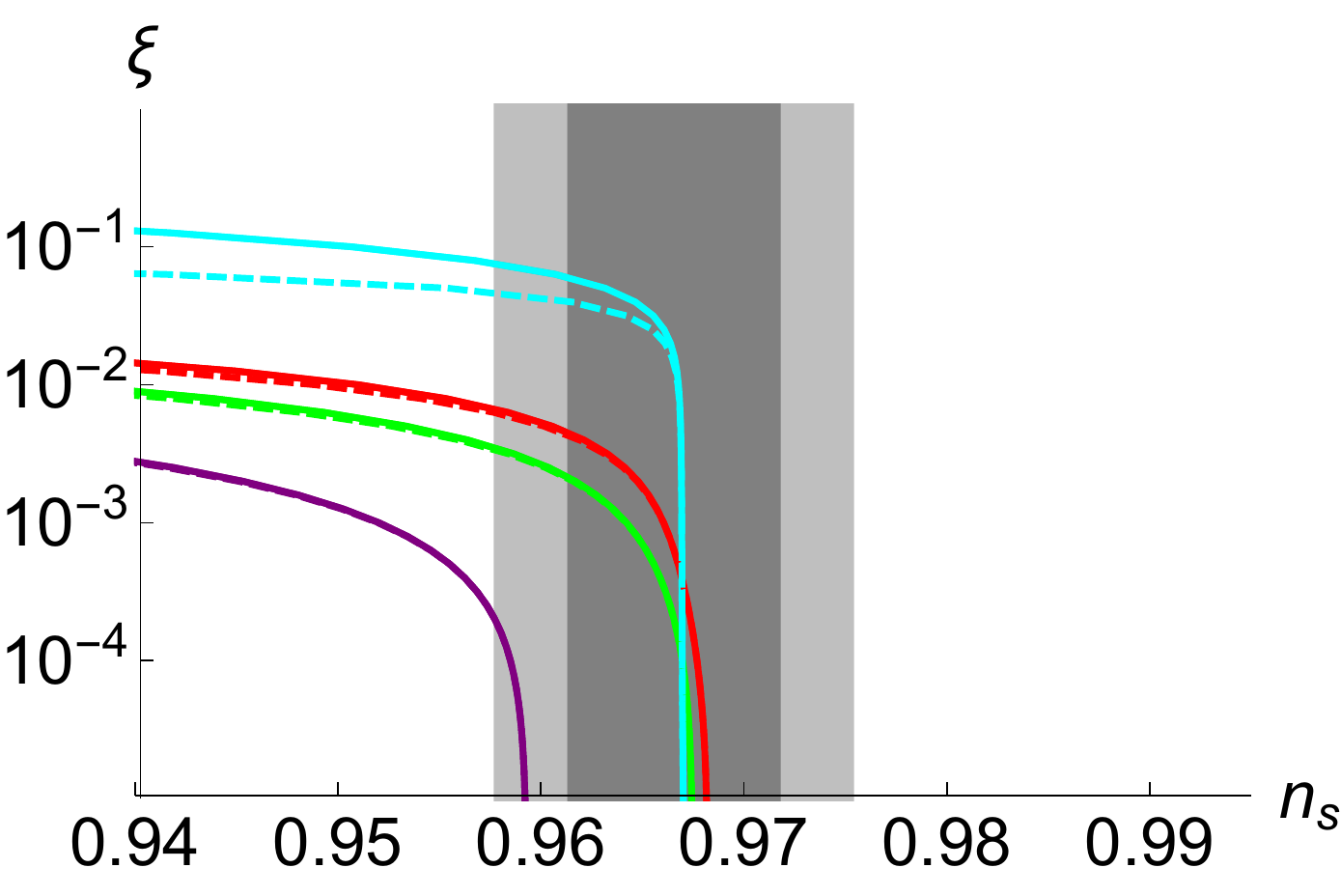}}%
    \subfloat[]{\includegraphics[width=0.46\textwidth]{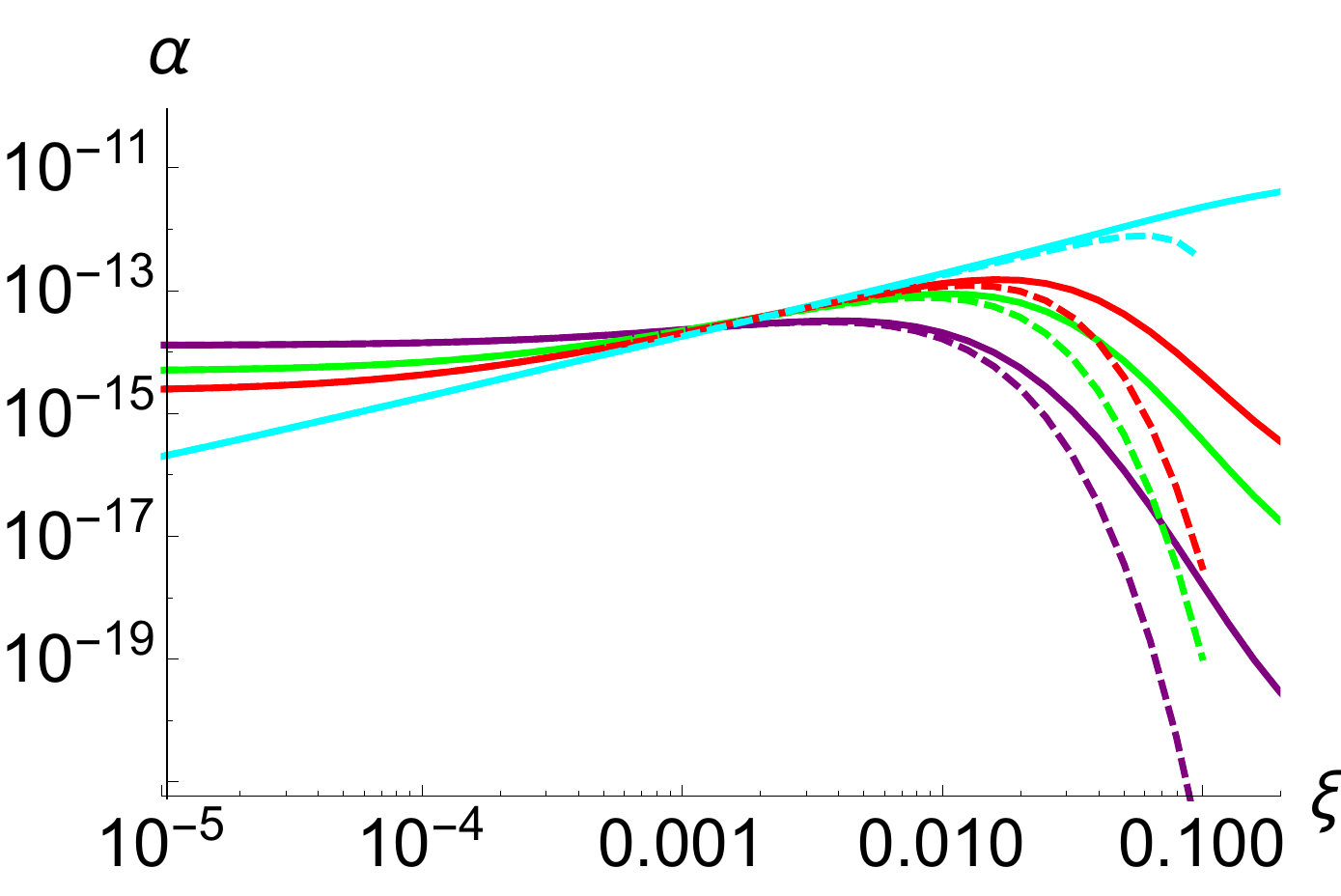}}%
    \caption{$r$ vs. $n_s$ (a), $r$ vs. $\xi$ (b), $\xi$ vs. $n_s$ (c) and $\alpha$ vs. $\xi$ (d) for $\delta=40,70,100,1000$ respectively in purple, green, red and cyan with $N_e =60$ $e$-folds in region 2. Continuous/dashed line represents metric/Palatini gravity. For reference we plot the predictions of CW inflation for $N_e=60$ (blue), $R^2$ (orange), $\chi^2$ (black) and $\chi^4$ (brown) inflation for $N_e \in [50,60]$. The gray areas represent the 1,2$\sigma$ allowed regions from Planck 2018 data~\cite{Planck2018:inflation}.}%
\label{fig:results:2:60}
\vskip 1cm
\begin{center}
    \includegraphics[width=0.46\textwidth]{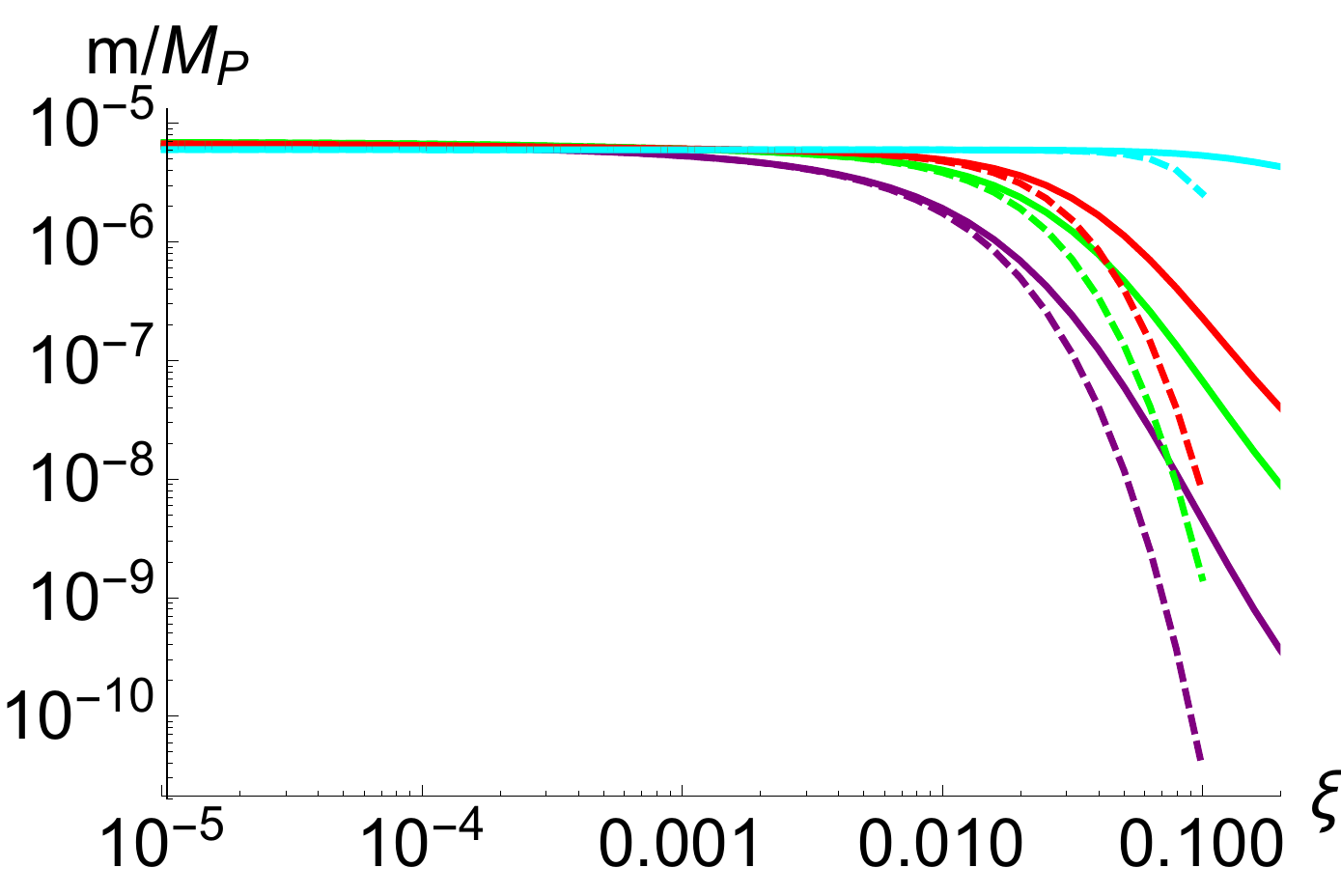}
\end{center}
\caption{$m$ vs. $\xi$ for $\delta=40,70,100,1000$ respectively in purple, green, red and cyan with $N_e =60$ $e$-folds in region 2. Continuous/dashed line represents metric/Palatini gravity.}
\label{fig:mvsxi:2:60}
\end{figure}

In this subsection we describe the phenomenology for inflation happening from the local maximum forward to second minimum, i.e. in the region $\phi_2 < \phi < \phi_3 $. Our results are presented in Figs. \ref{fig:results:2:50}-\ref{fig:mvsxi:2:60}. In Fig. \ref{fig:results:2:50} we show $r$ vs. $n_s$ (a), $r$ vs. $\xi$ (b), $\xi$ vs. $n_s$ (c) and $\alpha$ vs. $\xi$ (d) for $\delta=40,70,100,1000$ respectively in purple, green, red and cyan with $N_e =50$ $e$-folds. Continuous line represents metric gravity, while dashed line stands for Palatini gravity. For reference we plot the predictions of CW inflation for $N_e=50$ (blue), Starobinsky (orange), quadratic (black) and quartic (brown) inflation for $N_e \in [50,60]$. The gray areas represent the 1,2$\sigma$ allowed regions from Planck 2018 data~\cite{Planck2018:inflation}. Fig. \ref{fig:results:2:60} is the same as Fig. \ref{fig:results:2:50} but for $N_e=60$ $e$-folds. We notice again that the $r$ vs. $n_s$ results of metric and Palatini gravity are quite similar because the difference in the field redefinitions \eqref{eq:dphim} and \eqref{eq:dphiP} does not play a relevant role in this specific plot. In this case in the plotted region we have $\xi \lesssim 1$. However it is still possible to appreciate some difference in the $r$ vs. $\xi$, $\xi$ vs. $n_s$ and particularly in $\alpha$ vs. $\xi$ plots for $\xi \gtrsim 0.01$. At a given $\delta$ the non-minimal coupling first drives $r$ towards larger values with $\xi$ increasing and then towards smaller values as usual. On the other hand $n_s$ is driven towards lower values even more far away from the allowed Planck region. 
 Since $M=\delta M_P$ takes trans-Planckian values, we should check the value of inflaton mass. We can compute it exactly, obtaining
 \begin{equation}
  m_g = \delta  M_P \sqrt{\frac{2 \alpha }{1+ \xi  (1+6\xi) \delta^2}} \, , \label{eq:m:g}     
 \end{equation}
  \begin{equation}
  m_\Gamma = \delta  M_P \sqrt{\frac{2 \alpha }{1+ \xi \delta^2}} \, , \label{eq:m:gamma}
 \end{equation}
 where $m_g$ is the inflaton mass in metric gravity and $m_\Gamma$ is the inflaton mass in Palatini gravity. The corresponding results are shown in Figs. \ref{fig:mvsxi:2:50} and \ref{fig:mvsxi:2:60}. where we plot $m$ vs. $\xi$ with $\delta=40,70,100,1000$ respectively in purple, green, red and cyan for respectively $N_e =50,60$ $e$-folds. Continuous line represents metric gravity while dashed line stands for the Palatini case. We can see that the inflaton mass always remain sub-Planckian in both gravity formulations.
As final remark, we notice that, as usual, the difference in the number of $e$-folds does not affect the general behaviour of the results but only their eventual agreement with the observational constraints. For the chosen $\delta$ values, no line is within the allowed region for both $N_e=50, 60$ $e$-folds. However we can safely guess that for $N_e=60$ and $40 < \delta < 70$, the results are within the 2$\sigma$ region for a relatively small $\xi$.

\subsection{Large field inflation: $\phi > \phi_3$}
\begin{figure}[p]%
    \centering
    \subfloat[]{\includegraphics[width=0.46\textwidth]{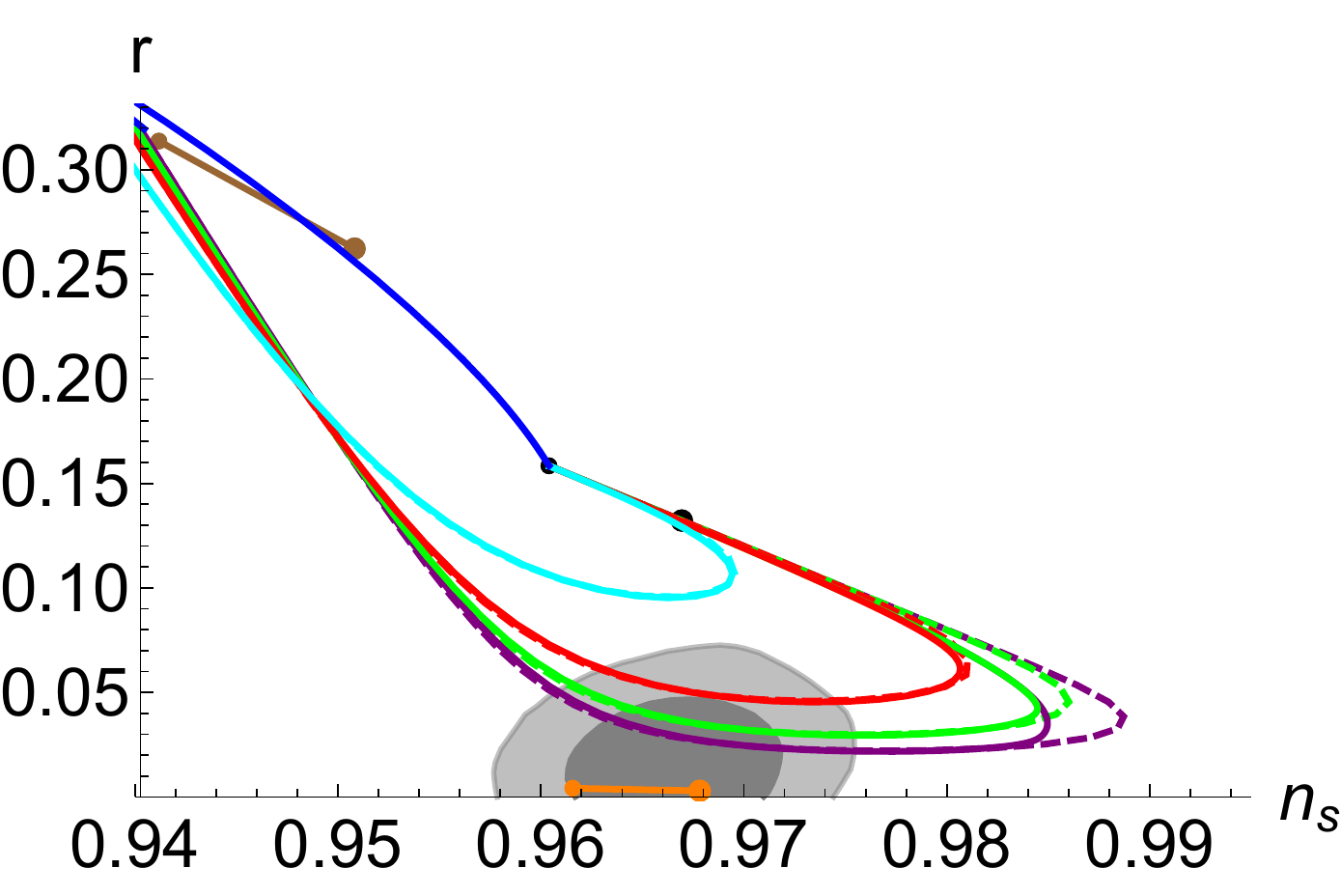}}%
    \subfloat[]{\includegraphics[width=0.46\textwidth]{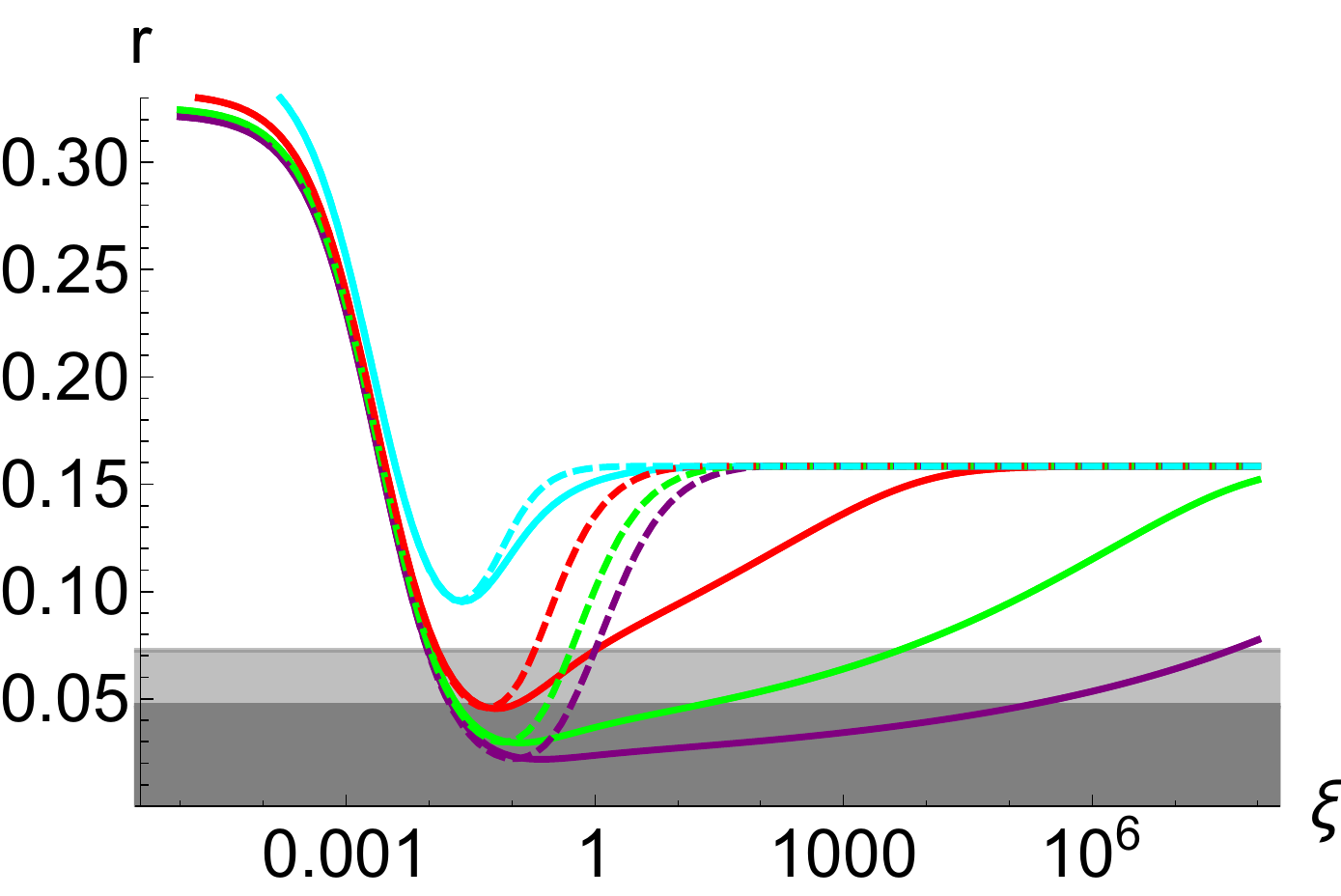}}\\
    \subfloat[]{\includegraphics[width=0.46\textwidth]{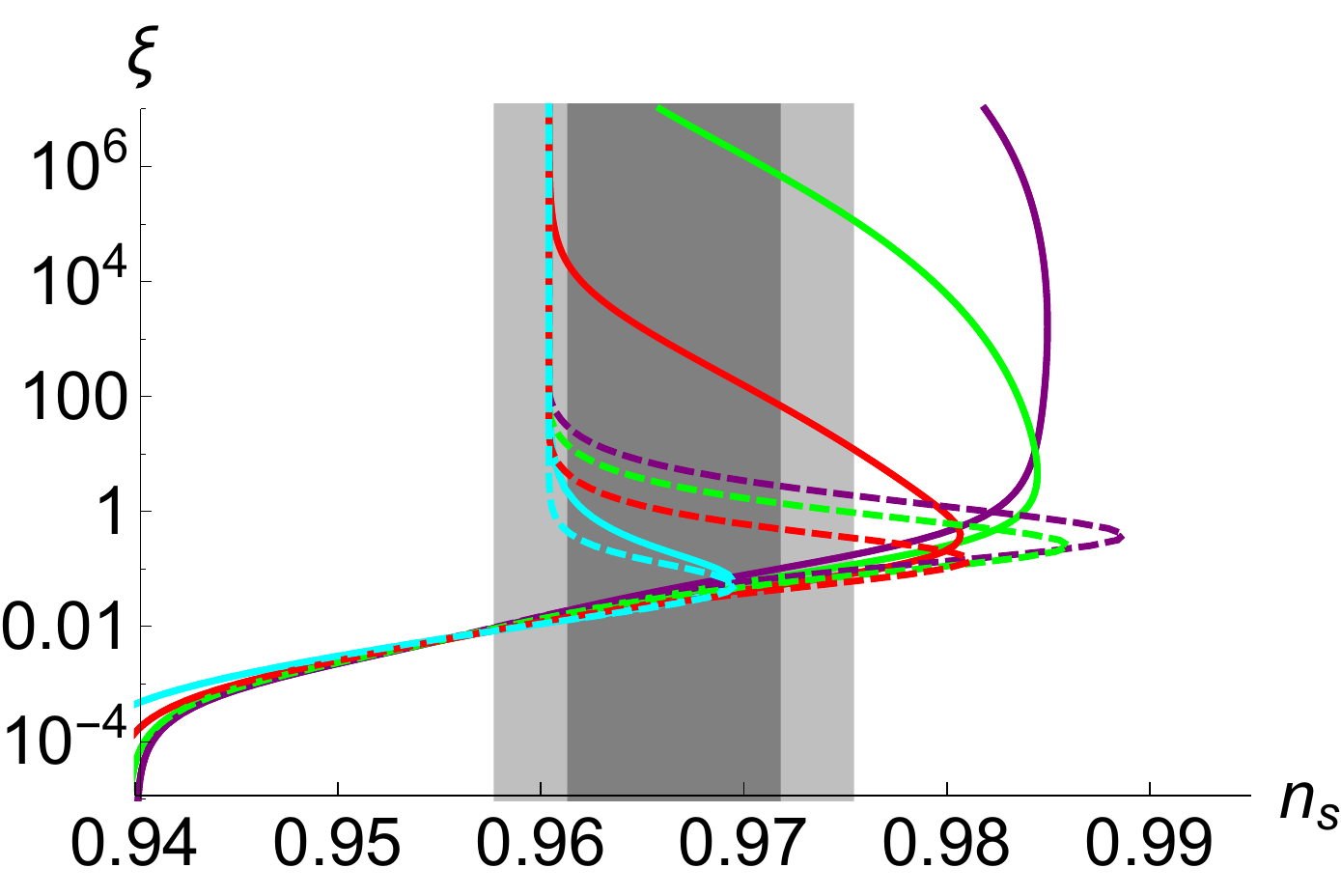}}%
    \subfloat[]{\includegraphics[width=0.46\textwidth]{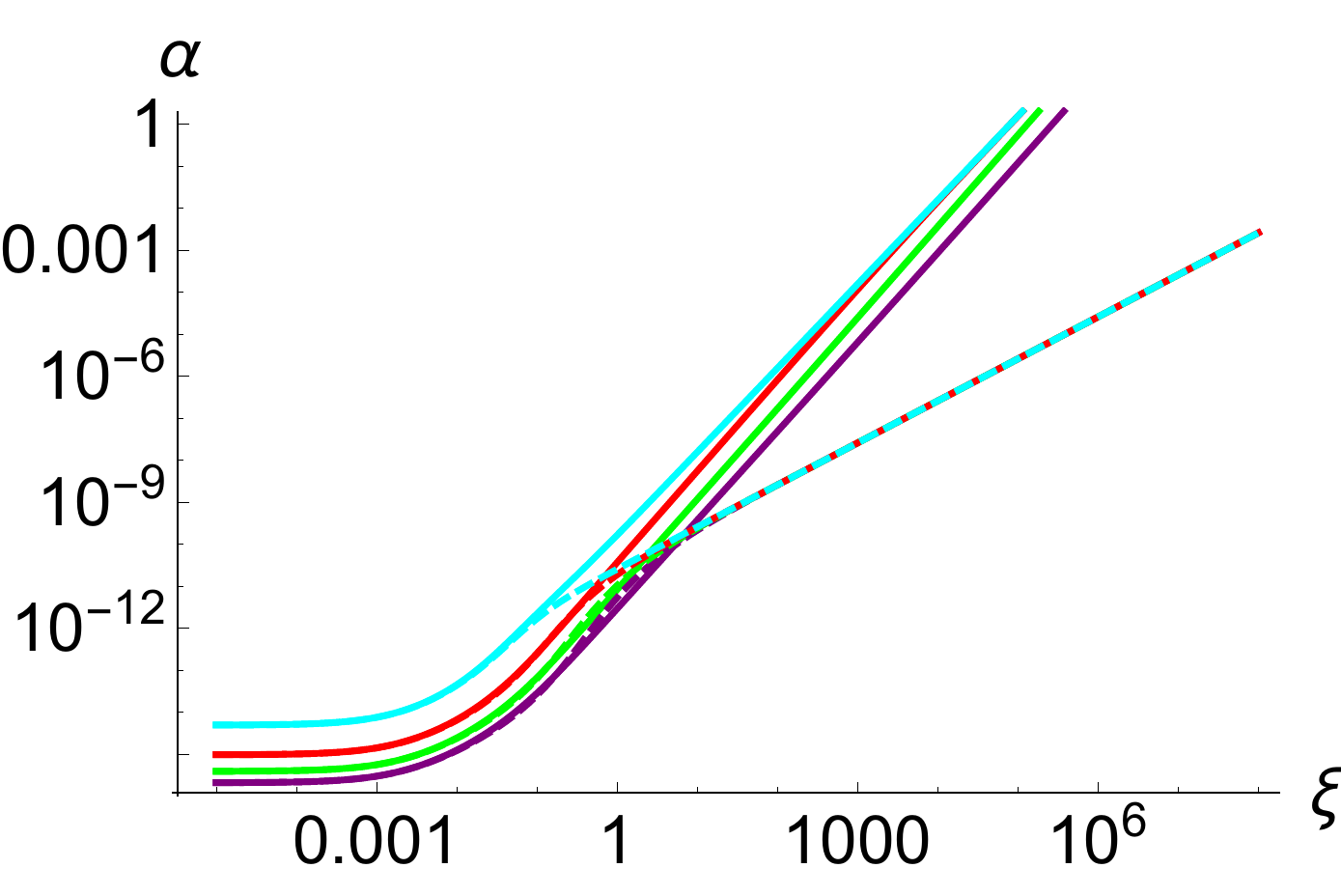}}%
    \caption{$r$ vs. $n_s$ (a), $r$ vs. $\xi$ (b), $\xi$ vs. $n_s$ (c) and $\alpha$ vs. $\xi$ (d) for $\delta=10^{-6},10^{-4},10^{-2},1$ respectively in purple, green, red and cyan with $N_e =50$ $e$-folds in region 3. Continuous/dashed line represents metric/Palatini gravity. For reference we plot the predictions of CW inflation for $N_e=50$ (blue), $R^2$ (orange), $\chi^2$ (black) and $\chi^4$ (brown) inflation for $N_e \in [50,60]$. The gray areas represent the 1,2$\sigma$ allowed regions from Planck 2018 data~\cite{Planck2018:inflation}.}%
\label{fig:results:3:50}
\vskip 1cm
\begin{center}
    \includegraphics[width=0.46\textwidth]{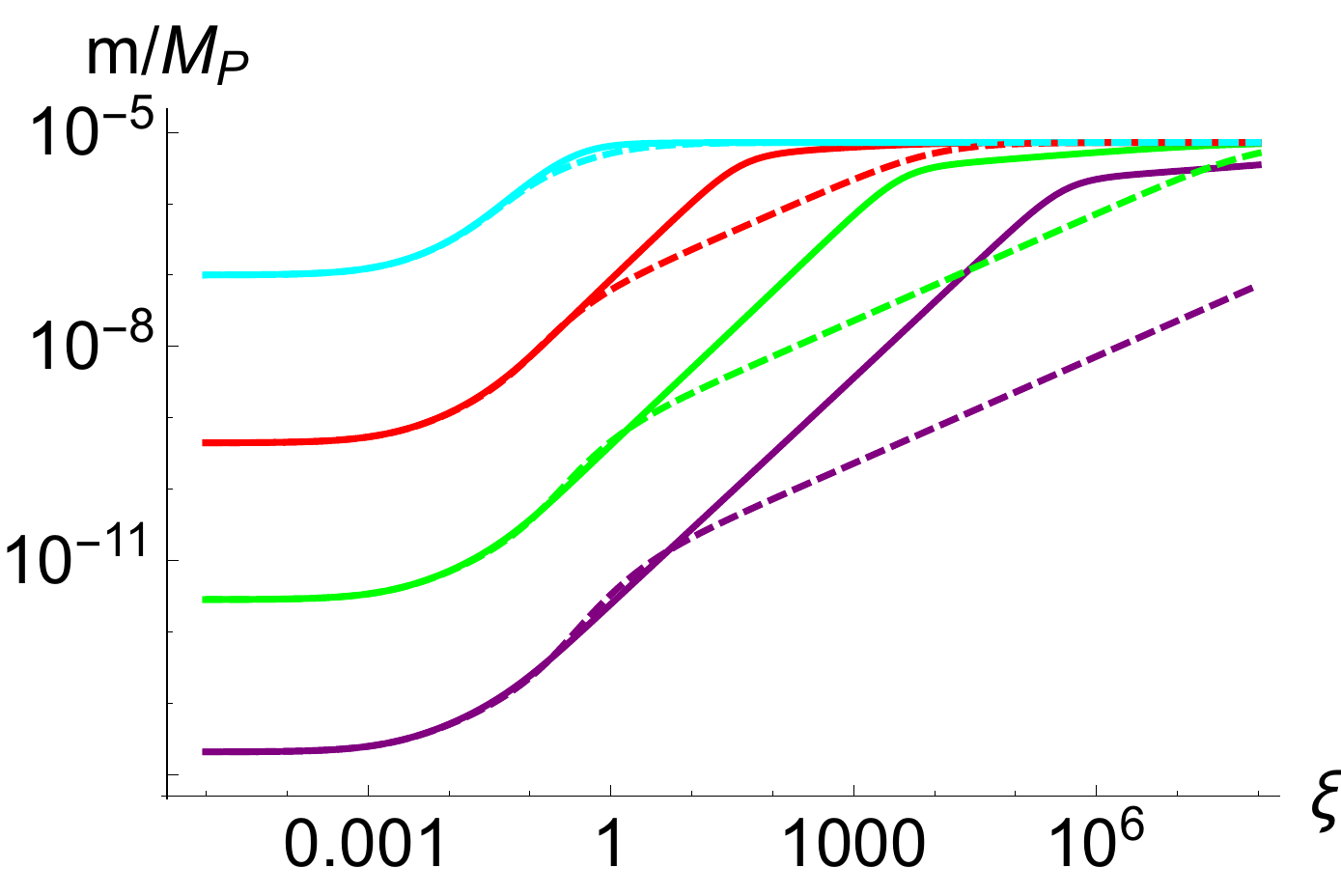}
\end{center}
\caption{$m$ vs. $\xi$ for $\delta=10^{-6},10^{-4},10^{-2},1$ respectively in purple, green, red and cyan with $N_e =50$ $e$-folds in region 3. Continuous/dashed line represents metric/Palatini gravity.}
\label{fig:mvsxi:3:50}
\end{figure}

\begin{figure}[p]%
    \centering
    \subfloat[]{\includegraphics[width=0.46\textwidth]{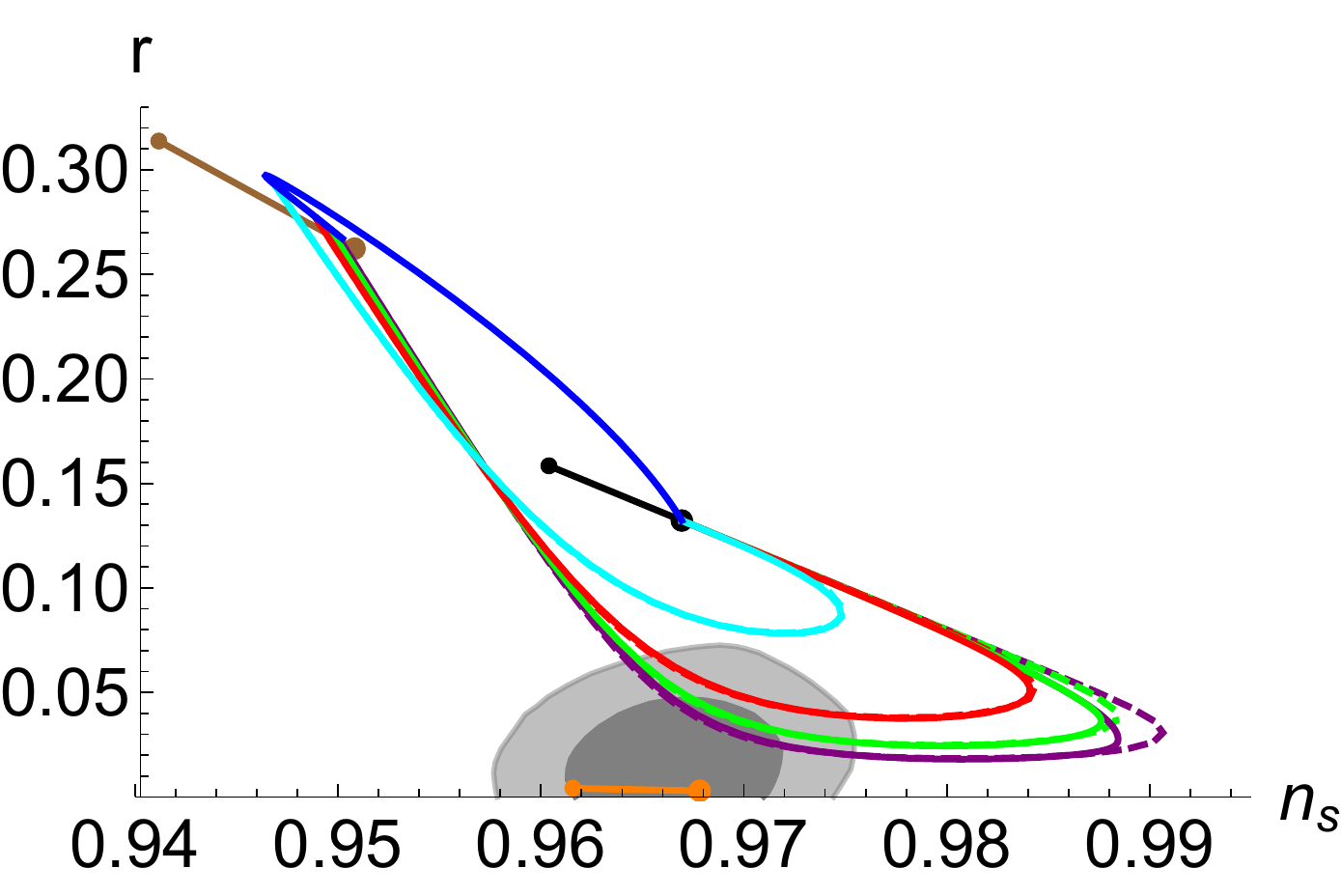}}%
    \subfloat[]{\includegraphics[width=0.46\textwidth]{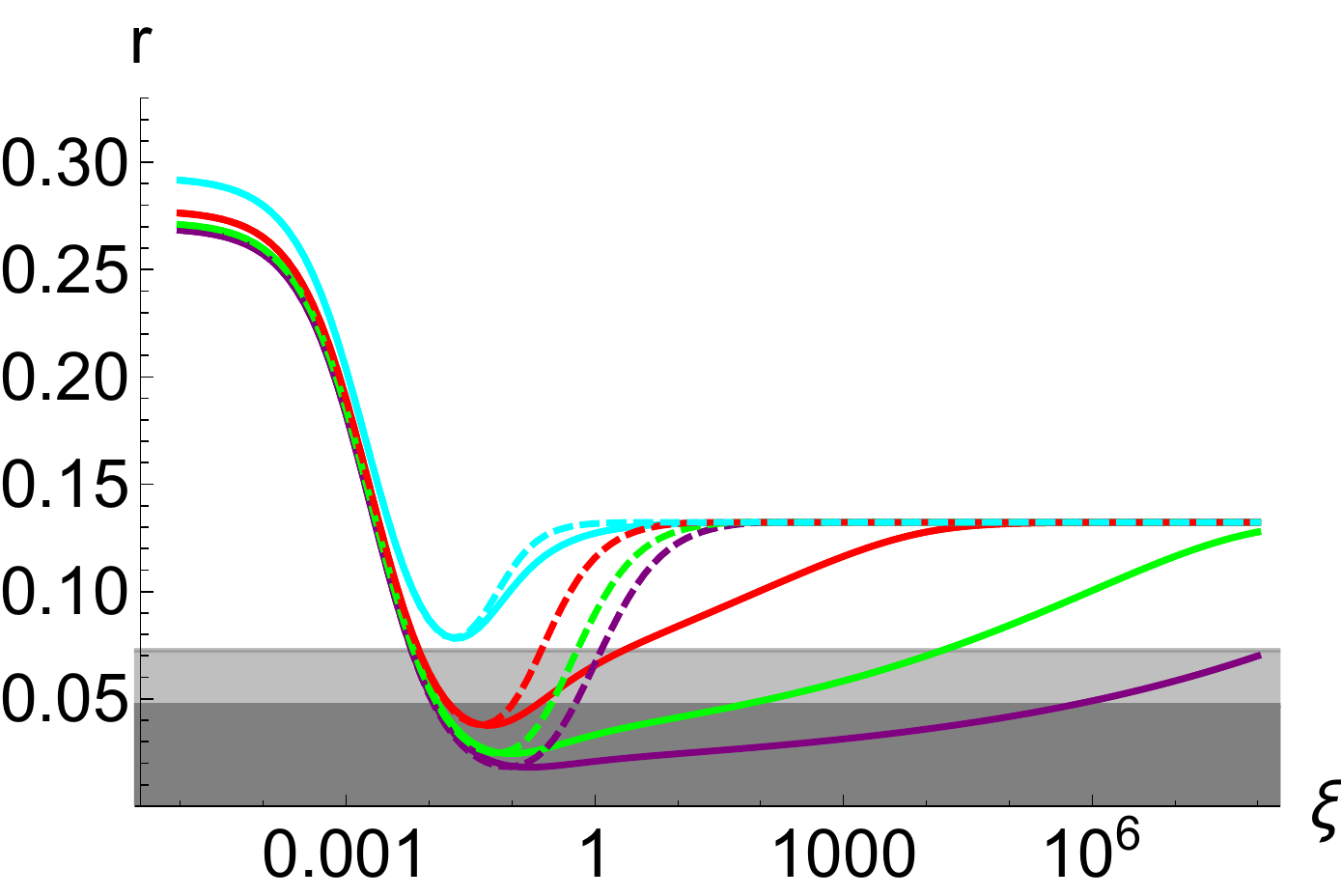}}\\
    \subfloat[]{\includegraphics[width=0.46\textwidth]{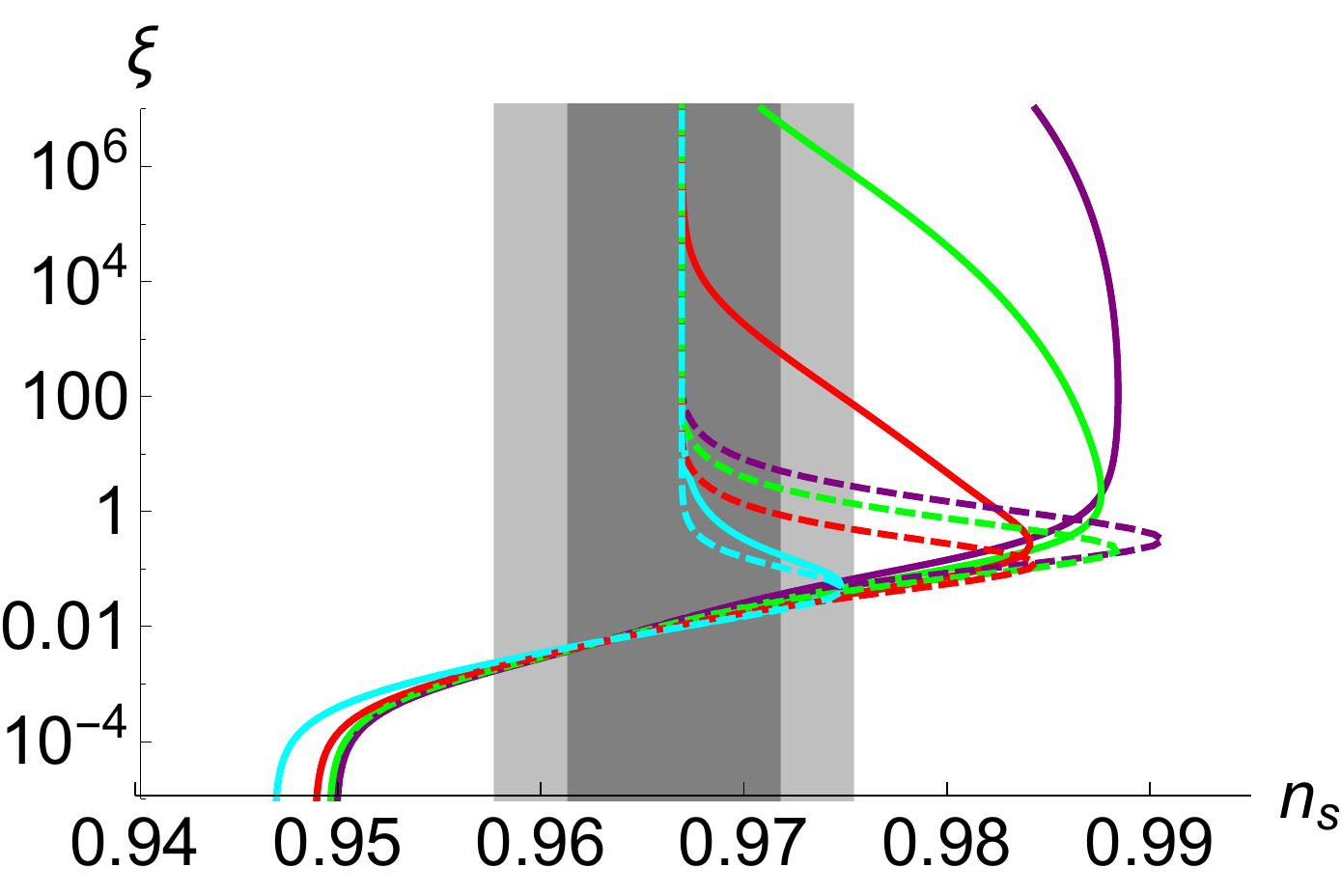}}%
    \subfloat[]{\includegraphics[width=0.46\textwidth]{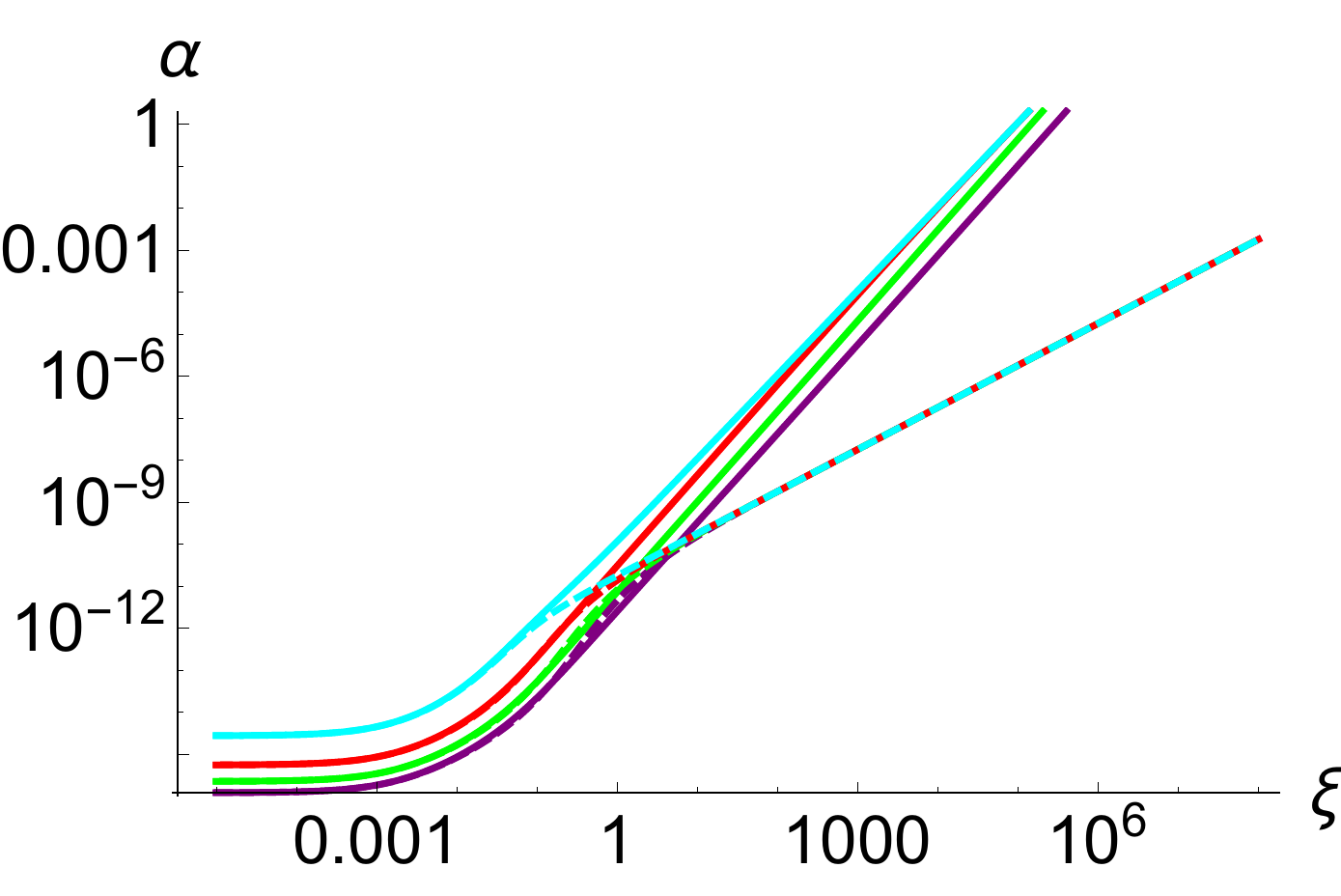}}%
    \caption{$r$ vs. $n_s$ (a), $r$ vs. $\xi$ (b), $\xi$ vs. $n_s$ (c) and $\alpha$ vs. $\xi$ (d) for $\delta=10^{-6},10^{-4},10^{-2},1$ respectively in purple, green, red and cyan with $N_e =60$ $e$-folds in region 3. Continuous/dashed line represents metric/Palatini gravity. For reference we plot the predictions of CW inflation for $N_e=60$ (blue), $R^2$ (orange), $\chi^2$ (black) and $\chi^4$ (brown) inflation for $N_e \in [50,60]$. The gray areas represent the 1,2$\sigma$ allowed regions from Planck 2018 data~\cite{Planck2018:inflation}.}%
\label{fig:results:3:60}
\vskip 1cm
\begin{center}
    \includegraphics[width=0.46\textwidth]{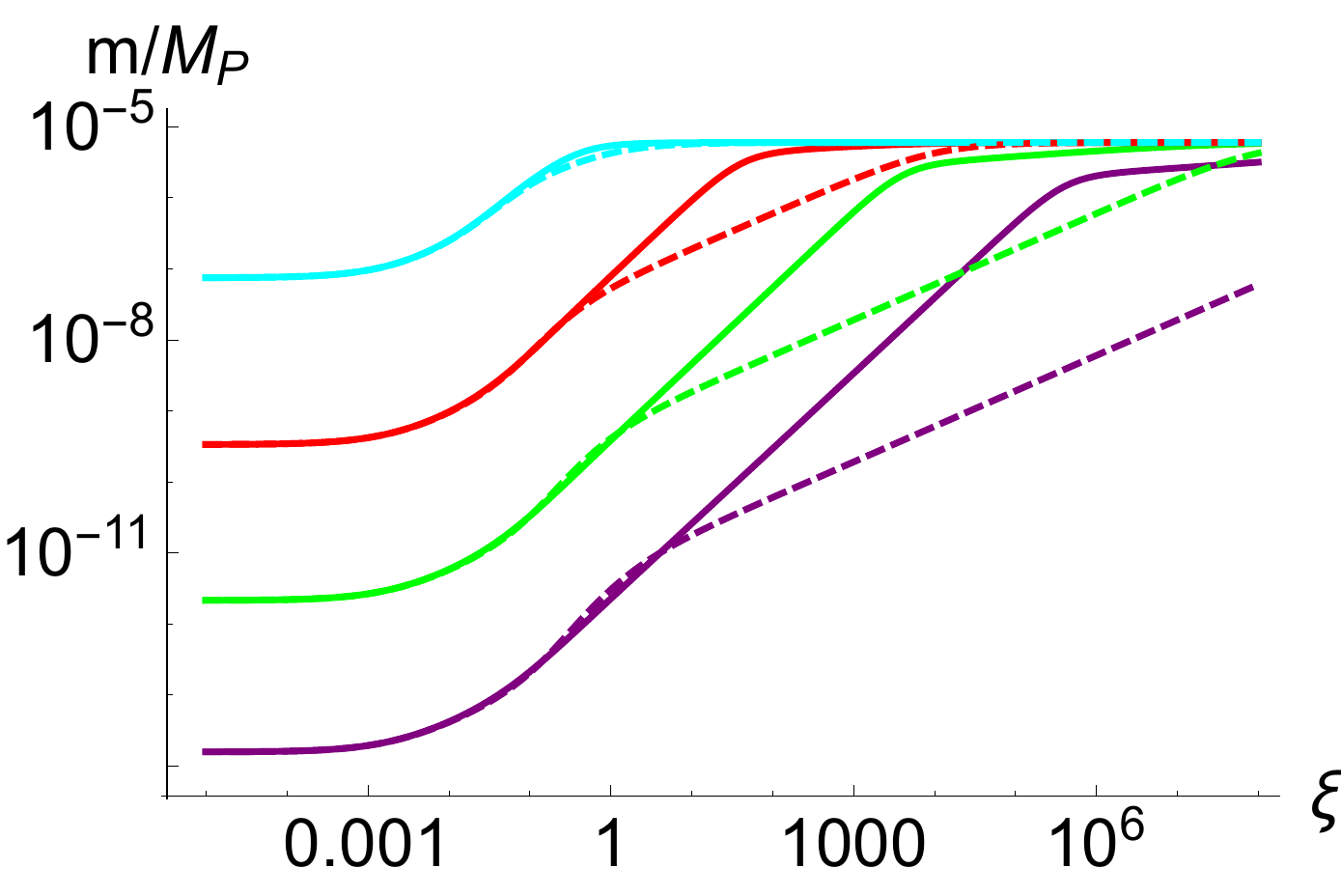}
\end{center}
\caption{$m$ vs. $\xi$ for $\delta=10^{-6},10^{-4},10^{-2},1$ respectively in purple, green, red and cyan with $N_e =60$ $e$-folds in region 3. Continuous/dashed line represents metric/Palatini gravity.}
\label{fig:mvsxi:3:60}
\end{figure}

In this subsection we describe the phenomenology for inflation happening from ``infinity" backward to the second minimum, i.e. in the region $\phi > \phi_3$.
 Our results are presented in Figs. \ref{fig:results:3:50}-\ref{fig:mvsxi:3:60}. In Fig. \ref{fig:results:3:50} we show $r$ vs. $n_s$ (a), $r$ vs. $\xi$ (b), $\xi$ vs. $n_s$ (c) and $\alpha$ vs. $\xi$ (d) for $\delta=10^{-6},10^{-4},10^{-2},1$ respectively in purple, green, red and cyan with $N_e =50$ $e$-folds. Continuous line represents metric gravity, while dashed line stands for Palatini gravity. For reference we plot the predictions of CW inflation for $N_e=50$ (blue), Starobinsky (orange), quadratic (black) and quartic (brown) inflation for $N_e \in [50,60]$. The gray areas represent the 1,2$\sigma$ allowed regions from Planck 2018 data~\cite{Planck2018:inflation}. Fig. \ref{fig:results:3:60} is the same as Fig. \ref{fig:results:3:50} but for $N_e=60$ $e$-folds. In most of the parameters space, the $r$ vs. $n_s$ results of metric and Palatini gravity are quite similar, while the $r$ vs. $\xi$, $\xi$ vs. $n_s$ and $\alpha$ vs. $\xi$ plots are quite different.
 %
 %
 
 When $\xi$ is relatively small the difference in the field redefinitions \eqref{eq:dphim} and \eqref{eq:dphiP} does not play a relevant role. However, in this case, $\xi$ might be still big enough to make a huge difference with respect to the minimally coupled case. For $0.01 < \xi < 0.1$ the results end up in the Planck allowed region for $\delta=10^{-6},10^{-4},10^{-2}$ and perturbativity ($\alpha<1$) is always maintained in both formulations.
 On the other hand, when $\xi \gg 1$, both formulations predict results in agreement with quadratic inflation. This can be proven quite easily. Taking the limit $\xi \to \infty$, we obtain for the Einstein frame potential
 \beq
  U(\phi) \approx  \frac{\alpha  M_P^4}{\xi ^2} \ln ^2\left(\frac{\phi }{\delta  M_P}\right) \, ,
 \eeq
 and for the field redefinition
 \beq
  \phi_g(\chi) \approx\delta  M_P \, e^{\frac{\chi }{\sqrt{6} M_P}} \, ,
 \eeq
 in the metric formulation or 
  \beq
  \phi_\Gamma(\chi) \approx \delta  M_P e^{\frac{\chi  \sqrt{\xi }}{M_P}} \, , 
 \eeq
 in the Palatini formulation, where we used the boundary condition $\chi(M)=0$ in solving both eqs. \eqref{eq:dphim} and \eqref{eq:dphiP}. Therefore we obtain the following limit 
  \beq
  U(\phi)_i \approx  \zeta_i \frac{\alpha M_P^2}{\xi}  \chi ^2 \, , \label{eq:U:quad:limit}
 \eeq
 where in the metric formulation 
 \beq
 i=g, \qquad \zeta_g=1/(6 \xi) \, , \label{eq:alpha:m}
 \eeq
 while in the Palatini formulation
  \beq
 i=\Gamma, \qquad \zeta_\Gamma=1 \, . \label{eq:alpha:P}
 \eeq
Hence the Einstein frame potential for $\xi \gg 1$ behaves quadratically regardless of the Jordan frame gravity formulation. Moreover, combining eqs. \eqref{eq:U:quad:limit}, \eqref{eq:alpha:m}, \eqref{eq:alpha:P} and \eqref{eq:As} we obtain
\beq
  \alpha_i \approx \frac{12 \pi ^2  A_s}{ \left(2 N_e+1\right){}^2}  \frac{\xi}{\zeta_i} \, ,
 \eeq
that explains why in the Palatini case $\alpha$ grows slower with $\xi$ increasing and remains within the perturbative bound ($\alpha < 1$) up to $\xi$ values larger than in the metric case. We also notice that we have an intermediate region (not ``too big'' and not ``too small'' $\xi$ values) where it is actually possible to discriminate in between the predictions of the two formulations also in the $r$ vs. $n_s$ plots. Unfortunately those results are out of the allowed $2\sigma$ region. 
 
Again, we check the value of the inflaton mass. The analytical expressions are already given in eqs. \eqref{eq:m:g} and \eqref{eq:m:gamma} respectively for the metric and the Palatini formulation. The corresponding numerical results are shown in Figs. \ref{fig:mvsxi:3:50} and \ref{fig:mvsxi:3:60}. where we plot $m$ vs. $\xi$ with $\delta=10^{-6},10^{-4},10^{-2},1$ respectively in purple, green, red and cyan for respectively $N_e =50,60$ $e$-folds. Continuous line represents metric gravity while dashed line stands for the Palatini case. We can see that the inflaton mass always remains sub-Planckian in both gravity formulations.

We notice again that the difference in the number of $e$-folds does not affect the general behaviour of the results but only their eventual agreement with the observational constraints. We see that when $M$ is sub-Planckian ($\delta <1$) it is often possible to find a region within the Planck constraints. Moreover, the lower the $\delta$, the better the agreement with the constraints.
As a final remark we notice that, for inflation happening in region 3, the allowed parameters space is larger for $N_e=50$.

\section{Conclusions}
\label{sec:conclusions}
In this article we studied a model of quartic inflation where radiative corrections allow for a realization of the MPCP i.e. an inflaton potential with two degenerate minima, one of them being the origin and the other one labelled as $M = \delta M_P$. $\delta$ has been considered a free parameter of the model. Since the potential is a continuous function with two minima, it inevitably exhibits also a local maximum. Therefore, inflation can happen in three distinct regions: 1) from the local maximum backward to the origin, 2) from the local maximum forward to the second minimum, 3) from ``infinity'' backward to the second minimum. We studied the inflationary predictions in all the three regions for $N_e=50,60$ $e$-folds. We found that only a small set of the parameters space is within the Planck constraints only for inflation happening in region 2) and for $N_e=60$. 

Therefore, we then analyzed the same model in presence of a Higgs-inflation-like non-minimal coupling to gravity. We studied the predictions of two different formulations of gravity, metric or Palatini. In most of the parameters space, the predictions in the $r$ vs. $n_s$ plot were indistinguishable. This is because the gravity formulations differ only for the Einstein frame canonical field redefinition and in the 2$\sigma$ Planck allowed region the non-minimal coupling is never big enough to make such a difference appreciable in the $r$ vs. $n_s$ plot. Eventual differences are appreciable only in the actual values of $\alpha$ and $\xi$, but only out of the 2$\sigma$ Planck allowed region. It is again possible to identify the three inflationary regions mentioned before, regardless of the adopted gravity formulation. Inflation happening in region 1) was still excluded in both gravity formulations, while inflation happening in region 2) remained mostly disfavoured. The most promising was slow-roll in region 3): the inflationary predictions cross even the 1$\sigma$ allowed region. Such a scenario can be either confirmed or ruled out by the forthcoming experiments (e.g. Simons Observatory \cite{SimonsObservatory:2018koc}, PICO \cite{NASAPICO:2019thw}, CMB-S4 \cite{Abazajian:2019eic} and LITEBIRD \cite{LiteBIRD:2020khw}).

\section*{Note}
This article is partially based on the BSc theses of J. Rajasalu \cite{Rajasalu} and K. Selke \cite{Selke}.

\acknowledgments

The author thanks Kristjan Kannike for useful discussions.
This work was supported by the Estonian Research Council grants MOBTT5, MOBTT86, PRG1055
and by the EU through the European Regional Development Fund
CoE program TK133 ``The Dark Side of the Universe''. 

\bibliographystyle{JHEP}
\bibliography{references}

\end{document}